\documentclass[aps,prb,showpacs,twocolumn, floatfix]{revtex4}
\usepackage{graphicx}

\newcommand{\beq}{\begin{equation}}
\newcommand{\eeq}{\end{equation}}
\newcommand{\bea}{\begin{eqnarray}}
\newcommand{\eea}{\end{eqnarray}}

\newcommand{\RR}{{\bf R}}
\newcommand{\RRp}{{\bf R'}}
\newcommand{\PP}{{\mathcal{P}}}
\newcommand{\dd}{{\bf d}}
\newcommand{\rr}{{\bf r}}
\newcommand{\gcc}{{g/cm$^{3}$}}

\begin{document}

\title{FPEOS: A First-Principles Equation of State Table of Deuterium for 
Inertial Confinement Fusion Applications}

\author{ S. X. Hu$^{1,*}$, B. Militzer$^2$, V. N. Goncharov$^1$, S. Skupsky$^1$ }

\affiliation{1. Laboratory for Laser Energetics, University of Rochester,
250 E. River Road, Rochester, NY 14623\\
2. Department of Earth and Planetary Science and 
Department of Astronomy, University of California, Berkeley, CA 94720
}

\date{\today}

\begin{abstract}
Understanding and designing inertial confinement fusion (ICF)
implosions through radiation-hydrodynamics simulations rely on the
accurate knowledge of the equation of state (EOS) of the deuterium
and tritium fuels.  To minimize the drive energy for ignition, the
imploding shell of DT-fuel needs to be kept as cold as possible.
Such low-adiabat ICF implosions can access to coupled and degenerate
plasma conditions, in which the analytical or chemical EOS models
become inaccurate. Using the path integral Monte Carlo (PIMC)
simulations we have derived a first-principles EOS (FPEOS) table
of deuterium that covers typical ICF fuel conditions at densities
ranging from 0.002 to 1596 {\gcc} and temperatures of 1.35
eV $-$ 5.5 keV. We report the internal energy and the pressure, and
discuss the structure of the plasma in terms of pair correlation
functions. When compared with the widely used {\sl SESAME} table and
the revised {\sl Kerley03} table, discrepancies in the internal
energy and in the pressure are identified for moderately coupled and
degenerate plasma conditions. In contrast to the {\sl SESAME} table,
the revised {\sl Kerley03} table is in better agreement with our
FPEOS results over a wide range of densities and temperatures.
Although subtle differences still exist for lower temperatures ($T <
10$ eV) and moderate densities ($1 - 10$ \gcc),
hydrodynamics simulations of cryogenic ICF implosions using the
FPEOS table and the {\sl Kerley03} table have resulted in similar
results for the peak density, areal density $\rho R$, and neutron
yield, which are significantly different from the {\sl SESAME}
simulations.
\end{abstract}

\pacs{52.25.Kn, 51.30.+i, 62.50.-p, 64.10.+h}

\maketitle

\section{Introduction}

Inertial confinement fusion (ICF) has been pursued for decades since
the concept was introduced in 1972 \cite{ICF}. In the traditional
central-hot-spot ignition designs, a capsule of cryogenic
deuterium-tritium (DT) covered with plastic ablator is driven to
implode either directly by intense laser pulses \cite{DDI} or
indirectly by x-rays in a hohlraum \cite{IDI}. To minimize the driving
energy required for ignition, the imploding DT-capsule needs to be
maintained as cold as possible \cite{Betti} for high compressions
(larger than a thousand times that of the solid DT density) at the
stagnation stage. This can either be done with fine-tuned shocks
\cite{GoncharovPRL_2010} or with ramp compression
waves. The reduction in temperature leads to pressures in the
imploding DT-shell that are just above the Fermi degeneracy pressure.
This is conventionally characterized by the so-called {\it adiabat}
parameter $\alpha = P/P_{F}$. Low-adiabat ICF designs with $1 <
\alpha < 2$ are currently studied with indirect-drive implosions at
the National Ignition Facility (NIF) \cite{NIF}. Direct-drive ignition
designs \cite{GoncharovPRL_2010} for NIF also place the DT-shell
adiabat at a low value of $ 2 < \alpha < 3$. Cryogenic DT targets
scaled from the hydro-equivalent NIF designs are routinely imploded
with a direct drive at the Omega laser facility \cite{omega}.

Since the compressibility of a material is determined by its equation
of state (EOS) \cite{Hu_PRL_2008}, the accurate knowledge of the EOS
of the DT-fuel is essential for designing ICF ignition targets and
predicting the performance of the target during ICF implosions.  To
perform radiation-hydrodynamics simulations of ICF implosions, one
needs to know the pressure and energy of the DT-fuel and the ablator
materials at various density and temperature conditions, which are
usually provided by EOS tables or analytical formulas. 
Various EOS tables for deuterium have
been assembled because its importance in ICF applications, planetary
science and high pressure physics. 

The widely used {\sl SESAME} EOS table of
deuterium~\cite{Kerley1972,Kerley2003} was based upon a {\it chemical
  model} for hydrogen~\cite{SaumonPRA1992, Ross1998, Rogers2001,
  Juranek2002} that describes the material in terms of well-defined
chemical species like H$_2$ molecules, H atoms and free protons and
electrons. Their interaction as well as many-body and degeneracy
effects are treated approximately. For the {\sl SESAME} table, liquid
perturbation theory was adopted in the molecular/atomic fluid phase
for ICF plasma conditions. A first-order expansion that only
takes into account nearest neighbor interactions was used in the
original {\sl SESAME} table~\cite{Kerley1972}. 

Chemical models are expected to work well in the regime of weak
coupling. However, in ICF implosions, the DT shell goes through a wide
range of densities from 0.1 up to 1000 {\gcc} and temperatures
varying from a few electron volts (eV) to several hundreds of electron
volts \cite{DDI, IDI}, which include plasma conditions with moderately
strong coupling. This provides the primary motivation for this paper,
where we derive the deuterium from first-principles path integral
Monte Carlo simulations~\cite{PierleoniPRL1994, MagroPRL1996, 
MilitzerPRL2000, MilitzerPRL2001}.

\begin{figure}
\rotatebox{0}{\includegraphics[scale=1.0]{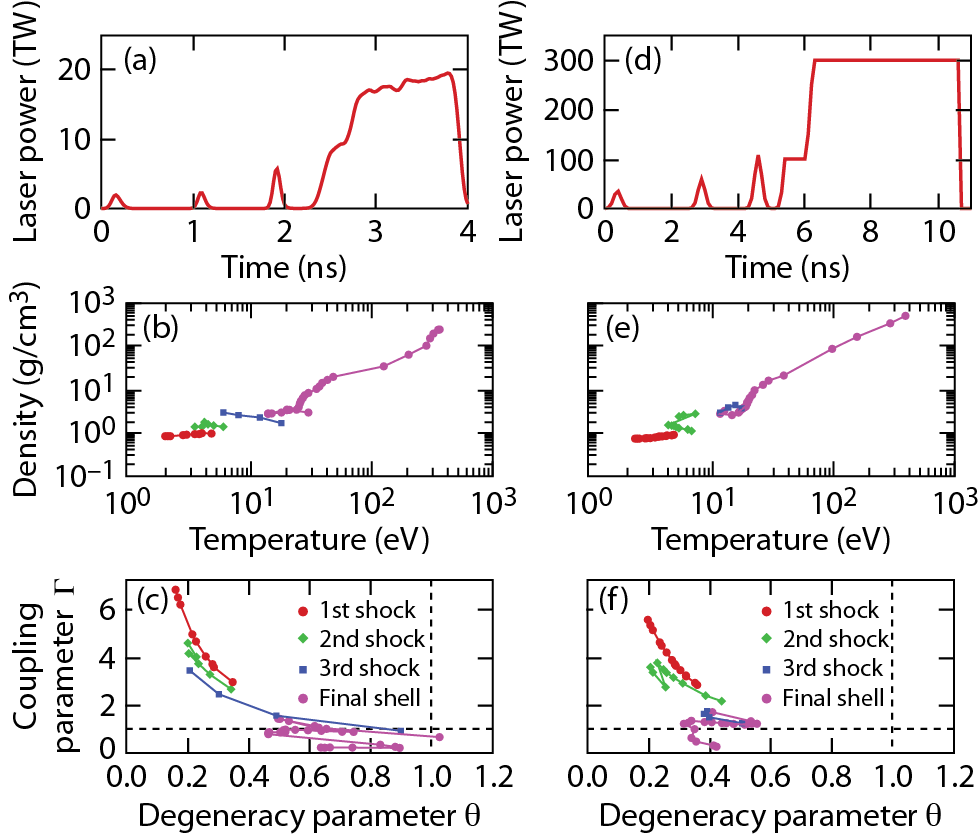}}
\caption{(Color online) (a)-(c): A cryogenic DT-implosion on OMEGA with the triple-picket step-pulse; 
(d)-(f): A direct-drive ignition 
design for NIF, scaled from hydro-equivalent OMEGA implosions.  
In both cases, strongly coupled and degenerated 
plasma conditions are indeed accessed.}
\label{conditions}
\end{figure}

The conditions for a low-adiabat ($\alpha \simeq 2.5$) cryogenic DT
implosion on OMEGA are shown in Fig.~\ref{conditions} on panels
(a)-(c). Panels (d)-(f) characterize the conditions for a direct-drive
ignition design for NIF that is hydro-equivalent to the OMEGA
implosion. In panels (a) and (d), we plot the laser pulse shapes.
Panels (b) and (e) show the density, $\rho$, and temperature, $T$,
path of the driven DT-shell that we derived with one dimensional (1D)
hydro-simulations using the hydro-code LILAC~\cite{LILAC}. The DT
shell is predicted to undergo a variety of drive stages including 
several shocks and the final push by the main pulse.

The $\rho$-$T$ path of the imploding DT shell can be projected onto a
plane spanned by the coupling parameter $\Gamma=1/(r_s k_b T)$ and the
degeneracy parameter $\theta=T/T_F$. $T_F=\frac{\hbar ^2}{2 m_e
  k_b}\times (3\pi^2 n)^{2/3}$ is the Fermi temperature of the
electrons in a fully ionized plasma and $r_s$ is the Wigner-Seitz
radius that is related to the number density of the electrons, $n=3/(4
\pi r_s^3)$.  One finds that the imploding shells indeed pass through
the strongly coupled ($\Gamma > 1$) and degenerate ($\theta < 1$)
regimes and one expects coupling and degeneracy effects to play
significant roles in the compression and yield-production in
low-adiabat ICF implosions~\cite{FPEOS_PRL}.
 
Strong coupling and degeneracy effects in ICF plasmas have recently
attracted much attention, as they may redefine the so-called
1D-physics of ICF implosions. The essential pieces of physics 
models used in ICF hydro-simulations, such as the electron-ion energy
relaxation rate \cite{relax-rates}, the thermal conductivity
\cite{French_PRL_2008}, the fusion reaction rate \cite{Polluk2004},
and viscosity and mutual diffusion in deuterium-tritium mixtures
\cite{Kress2010} in coupled and degenerated plasmas have been
re-examined recently with experimental and theoretical methods. EOS
measurements of liquid deuterium along the principal Hugoniot reaching
about 100-200 GPa have been performed using laser-driven shock
waves~\cite{Silva1997,Collins1998Science,Collins1998POP,MostovychPRL,Boehly2004,Hicks2009},
magnetically driven flyers~\cite{Knudson2001, Knudson2004PRB}, and
convergent explosives \cite{Belov2002, Fortov2007}. First-principles
computer simulations have emerged as the preferred theoretical tool to
derive the EOS of deuterium under such extreme conditions. Two methods
have been most successful: density functional molecular
dynamics (DFT-MD) \cite{LACollins1995, Lenosky2000, GalliPRB2000,
  CollinsPRB2001, Clerouin2001, DesjarlaisPRB2003, BonevPRB2004} and
the path integral Monte Carlo (PIMC)~\cite{PierleoniPRL1994,
  MagroPRL1996, MilitzerPRL2000, MilitzerPRL2001}.  In
contrast to chemical models, these first-principles methods can 
take many-body effects fully into account. Results from such simulations
have also been used to revise the original {\sl SESAME} EOS table of
deuterium to yield the improved {\sl Kerley03} EOS table~\cite{Kerley2003}.

For ICF applications, we are especially concerned about the EOS
accuracy along the implosion path in the density-temperature plane,
i.e., in the range of $\rho = 0.1 - 1000$ g/cm$^3$ and $T = 1 - 1000$
eV.  For such high temperatures, standard DFT methods become
prohibitively expansive because of the large number of electronic
orbitals that would need to be included in the calculation to account for
electronic excitations~\cite{Collins_QMD}. Orbital-free semi-classical
simulation methods based on Thomas-Fermi theory~\cite{OFMD} is more
efficient but they approximate electronic correlation effects and
cannot represent chemical bonds. Therefore, in current form, they
cannot describe the systems at lower temperatures accurately.

Path integral Monte Carlo has been shown to work rather well for EOS
calculations of low-Z materials such as deuterium \cite{FPEOS_PRL} and
helium~\cite{Mi06,Milizer_He_2009}.  In this paper we present a
first-principles equation of state (FPEOS) table of deuterium from
restricted PIMC simulations~\cite{CeperleyRMP1995}. This method has
been successfully applied to compute the deuterium EOS
\cite{MilitzerPRL2000,MiltzerPhD} up to a density of $\rho=5.388$
g/cm$^3$. At lower temperatures, the PIMC results have been shown to
agree well with DFT-MD calculations for
hydrogen~\cite{MilitzerPRL2001} and more recently for
helium~\cite{Milizer_He_2009}.

Our FPEOS table derived from PIMC covers the whole DT-shell plasma
conditions throughout the low-adiabat ICF implosions.  Specifically,
our table covers densities ranging from 0.002 to $1596$ g$\,$cm$^{-3}$
and temperatures of 1.35 eV $-$ 5.5 keV. When compared with the widely
used {\sl SESAME-EOS} table and the revised {\sl Kerley03-EOS} table,
discrepancies in the internal energy and the pressure have been
identified in moderately coupled and degenerate regimes.
Hydrodynamics simulations for cryogenic ICF implosions using our FPEOS
table and the {\sl Kerley03-EOS} table have resulted in similar peak
density, areal density $\rho R$, and neutron yield, which differ
significantly from the {\sl SESAME} simulations.

The paper is organized as follows. A brief description of the path
integral Monte Carlo method is given in Sec. II. In Sec. III our FPEOS
table is presented. In Sec. IV, we characterize the properties of the
deuterium plasma for a variety of density and temperature conditions in
terms of pair correlation functions.  Comparisons between the FPEOS
table, the {\sl SESAME} and the {\sl Kerley03} EOS as well as the
simple Debye-H\"uckel plasma model are made in Sec. V. In Sec. VI, we
analyze the implications of different EOS tables for ICF applications
through hydro-simulations and comparisons with experiments. The paper
is summarized in Sec. VII.

\section{The Path integral Monte Carlo method}

Path integral Monte Carlo (PIMC) is the appropriate computational
technique for simulating many-body quantum systems at finite
temperatures. In PIMC calculations, electrons and ions are treated on
equal footing as paths, which means the quantum effects of both
species are included consistently, although for the temperatures under
consideration, the zero-point motion and exchange effects of the
nuclei are negligible. 

The fundamental idea of the path integral approach is that the density
matrix of a quantum system at temperature, $T$, can be expressed as a
convolution of density matrices at a much higher temperature, $M \times
T$:
\begin{eqnarray}
\rho({\bf R,R'};\beta) &=& \int d{\bf R}_1 \, d{\bf R}_2 \, \cdots \, d{\bf R}_{M-1} \, 
\rho({\bf R,R}_1;\Delta \beta) \nonumber \\
  & \times& \rho({\bf R}_1,{\bf R}_2;\Delta \beta) \cdots \rho({\bf R}_{M-1},{\bf R'};\Delta \beta).
\label{PI}
\end{eqnarray}
This is an exact expression. The integral on the right can be
interpreted as a weighted average over all {\it paths} that connect
the points ${\bf R}$ and ${\bf R'}$. ${\bf R}$ is a collective
variable that denote the positions of all particles ${\bf R} =\{ {\bf
  r}_1, \ldots {\bf r}_N \}$. $\beta=1/k_bT$ represents length of the
path in ``imaginary time" and $\Delta \beta = \beta / M$ is the size of
each of the $M$ time steps. 

From the free particle density matrix which can be used 
for the high-temperature density-matrices, 
\begin{equation}
\label{single_gaussian}
\rho^{[1]}_0({\bf r, r'};\beta) = (2 \pi \hbar^2 \beta /m)^{-d/2} 
\exp \left \{ -\frac{({\bf r}-{\bf r}')^2}{2 \hbar^2 \beta/m}
\right \}\;,
\end{equation}
one can estimate that the separation of two
adjacent positions on the path, $\Delta r = {\bf r}_{i+1} - {\bf
  r}_{i}$ can only be on the order of $\sqrt{\hbar^2 \Delta \beta /
  m}$ while the separation of the two end points is approximately
$\sqrt{\hbar^2 \beta / m}$. One can consequently interpret the
positions ${\bf R}_1 \ldots {\bf R}_{M-1}$ as intermediate points on a
path from ${\bf R}$ and ${\bf R'}$. The multi-dimensional integration
over all paths in Eq.~\ref{PI} can be performed efficiently with Monte
Carlo methods~\cite{CeperleyRMP1995}.

In general observables associated with operator, $\hat{O}$, can be derived from,
\beq
\left<\hat{O}\right> = \frac{
\int \dd \RR \int \dd \RRp \left< \RR \right| \hat{O} \left| \RRp \right> \rho(\RRp,\RR;\beta)}
{\int \dd \RR \; \rho(\RR,\RR;\beta)} \;,
\eeq
but for the kinetic and potential energies, $E_K$ and $E_P$, as well
as for pair correlation functions only diagonal matrix elements
($\RR=\RR'$) are needed. The total internal energy follows from $E=E_K+E_P$
and the pressure, $P$, can be obtained from the virial theorem for Coulomb
systems, 
\beq 
P = (2 E_K + E_P )/3V\;. 
\eeq 
$V$ is the volume.

Electrons are fermions and their fermionic characters matters for the
degenerate plasma conditions under consideration. This implies one
needs to construct an antisymmetric many-body density matrix, which
can be derived by introducing a sum of all permutations, $\PP$, and
then also include paths from $\RR$ to $\PP \RR'$. While this approach
works well for bosons~\cite{CeperleyRMP1995}, for fermions each
permutation must be weighted by a factor $(-1)^\PP$. The partial
cancellation of contributions with opposite signs leads to an
extremely inefficient algorithm when the combined position and
permutation space is sampled directly. This is known as {\it Fermion
  sign problem}, and its severity increases as the plasma becomes more
degenerate.

We deal with the Fermion sign problem by introducing the fixed node
approximation~\cite{Ce91,Ce96},
\begin{equation}
\rho_F(\RR, \RR' ;\beta) =
\frac{1}{N!}\; \sum_\PP \; (-1)^\PP 
\! \! \! \! \! \! \! \! \! \! \! \! \! 
\int\limits_{
\begin{array}{c}
\scriptstyle
\RR \rightarrow \PP \RR' \\[-1mm]
\scriptstyle
\rho_{\scriptscriptstyle T}(\RR,\RR_t; t)>0
\end{array}
}
\! \! \! \! \! \! \! \! \! \! \! \! \! 
\dd\RR_t \;\; e^{-S[\RR_t] },
\label{restricted_PI} 
\end{equation}
where one only includes those paths that satisfy the nodal constraint,
$\rho_{T}(\RR,\RR_t; t)>0$, at every point. $S[\RR_t]$ is
the action of the path and $\rho_T$ is a fermionic trial density a
matrix that must be given in analytic form. For this paper, we rely on free particle nodes,
\beq
\rho_T(\RR,\RRp;\beta)=\left|
\begin{array}{ccc}
\rho^{[1]}(\rr_{1},\rr'_{1};\beta)&\ldots&\rho^{[1]}(\rr_{N},\rr'_{1};\beta)\\
\ldots&\ldots&\ldots\\
\rho^{[1]}(\rr_{1},\rr'_{N};\beta)&\ldots&\rho^{[1]}(\rr_{N},\rr'_{N};\beta)
\end{array}\right|. \label{matrixansatz}
\eeq 
but the nodes of a variational density matrix~\cite{MP00} have
also been employed in PIMC
computations~\cite{MilitzerPRL2000,Mi06,Milizer_He_2009}.

We have performed a number of convergence tests to minimize errors
from using a finite time step and from a finite number of particles in
cubic simulation cells with periodic boundary conditions. We
determined a time step of $\Delta \beta \le \left[ 100 \times k_b T_F
\right]^{-1}$ was sufficient to accurately account for all
interactions and degeneracy effects. We perform our PIMC calculations with different 
numbers of atoms depending on the deuterium density:
$N=64$ atoms for $\rho < 2.5$ g$\,$cm$^{-3}$, $N=128$ atoms for $2.5 <\rho < 10.5$ g$\,$cm$^{-3}$.
and $N=256$ atoms for $\rho > 10.5$ g$\,$cm$^{-3}$.

\section{The FPEOS table of deuterium}

\begin{figure}
\rotatebox{0}{\includegraphics[scale=1.0]{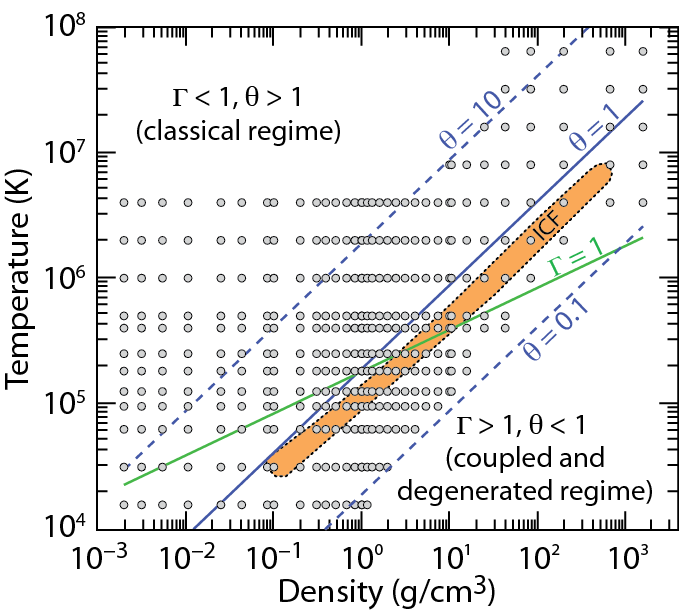}}
  \caption{(Color online) The temperature and density conditions
    covered by the FPEOS table. The gray circles represent our PIMC
    calculations, while the shell conditions in ICF implosions are
    schematically shown by the region in orange color. The blue and
    green lines of $\theta =1$ and $\Gamma =1$ characterize the
    boundaries of degeneracy and coupling conditions, respectively.  }
\label{dT-range}
\end{figure}

We have carried out PIMC calculations for a variety of density and
temperature conditions that are of interests to inertial confinement
fusion applications. The resulting FPEOS table for deuterium covers
the density range from 0.0019636 g$\,$cm$^{-3}$ ($r_s=14$ in units of Bohr radii $a_0$) to
1596.48802 g$\,$cm$^{-3}$ ($r_s=0.15$ $a_0$) and the temperature interval
from 15$\,$625 K ($\simeq 1.35$ eV) to $6.4\times 10^7$ K ($\simeq
5515.09$ eV). Fig.~\ref{dT-range} shows the conditions for every
simulation combined with lines for $\Gamma =1$ and $\theta=1$ to
indicate the boundaries between coupled/uncoupled and
degenerate/non-degenerate plasma conditions. Plasma conditions in the
upper left corner of the diagram are weakly coupled and classical
($\Gamma \ll 1$ and $\theta \gg 1$), while the lower right of the diagram
represent strongly coupled and highly degenerate conditions ($\Gamma
\gg 1$ and $\theta \ll 1$). The lowest temperatures in our PIMC
calculations reach to the regime of $\theta \simeq 0.1$. 

To give an example for the ICF plasma conditions, we added the
conditions of an imploding DT capsule shown in Fig.~\ref{conditions} to
Fig.~\ref{dT-range}. It can be seen that the DT shell undergoes a
change from strongly coupled to an uncoupled regime during the shock
transits. The electronic conditions change from fully degenerated to
partially degenerate accordingly.  All these conditions are covered
by our PIMC results that we have assembled into the following FPEOS
table~\ref{tab1}. The pressure and the internal energy as well as
their statistical error bars from PIMC simulations are listed for different
density and temperature conditions.

\begin{table}
 \caption{FPEOS table with pressures and internal energy per atom for 
deuterium. The statistical uncertainties from the PIMC simulations are 
in given in brackets e.g. 0.219(5)=0.219$\pm$0.005, 414.4(1.6)=414.4$\pm$1.6, 
or 70230(400)=70230$\pm$400.}
  \begin{tabular}{ccc}
  \toprule
  Temperature  &  Pressure  &  Internal energy \\
  (K)  &  (Mbar)  &  (eV$/$atom) \\
 \hline 
\multicolumn{3}{c}{ $\rho$ = 1.96360$\times 10^{-3}$g$\,$cm$^{-3}$ [$r_s =14.0$~$a_0$]}\\
        15625&   0.001290(8) & -10.83(3)\\
        31250&   0.003364(8) & -2.867(16)\\
        62500&   0.009046(8) & 11.930(13)\\
        95250&   0.014640(8) & 21.870(12)\\
       125000&   0.019600(10) & 30.080(15)\\
       181825&   0.028920(8) & 45.190(13)\\
       250000&   0.040040(12) & 63.04(2)\\
       400000&   0.06440(2) & 102.00(4)\\
       500000&   0.08074(2) & 128.10(4)\\
      1000000&   0.16170(3) & 257.30(6)\\
      2000000&   0.32390(6) & 515.90(10)\\
      4000000&   0.64830(14) & 1033.0(2)\\
\multicolumn{3}{c}{$\rho$ = 3.11810$\times 10^{-3}$g$\,$cm$^{-3}$ [$r_s =12.0$~$a_0$]}\\
        15625&   0.002048(14) & -10.97(3)\\
        31250&   0.005105(14) & -3.82(2)\\
        62500&   0.013980(12) & 10.860(12)\\
        95250&   0.022970(11) & 21.180(12)\\
       125000&   0.030840(15) & 29.490(15)\\
       181825&   0.045690(13) & 44.720(13)\\
       250000&   0.063400(19) & 62.67(2)\\
       400000&   0.10220(4) & 101.80(4)\\
       500000&   0.12790(3) & 127.70(4)\\
      1000000&   0.25670(5) & 257.10(5)\\
      2000000&   0.51440(10) & 516.00(10)\\
      4000000&   1.0290(2) & 1033.0(2)\\
\multicolumn{3}{c}{ $\rho$ = 5.38815$\times 10^{-3}$g$\,$cm$^{-3}$ [$r_s =10.0$~$a_0$]}\\
        15625&   0.00349(2) & -11.280(20)\\
        31250&   0.00845(2) & -4.841(18)\\
        62500&   0.02325(2) & 9.401(14)\\
        95250&   0.038870(19) & 20.080(12)\\
       125000&   0.05264(3) & 28.620(16)\\
       181825&   0.07841(2) & 44.040(13)\\
       250000&   0.10900(3) & 62.090(19)\\
       400000&   0.17630(7) & 101.40(4)\\
       500000&   0.22080(7) & 127.30(4)\\
      1000000&   0.44360(9) & 257.00(5)\\
      2000000&   0.88830(19) & 515.50(11)\\
      4000000&   1.7780(4) & 1033.0(2)\\
\multicolumn{3}{c}{ $\rho$ = 1.05237$\times 10^{-2}$g$\,$cm$^{-3}$ [$r_s =8.0$~$a_0$]}\\
        15625&   0.00659(5) & -11.550(17)\\
        31250&   0.01580(6) & -5.860(20)\\
        62500&   0.04325(4) & 7.477(13)\\
        95250&   0.07371(4) & 18.450(12)\\
       125000&   0.10080(5) & 27.220(16)\\
       181825&   0.15150(4) & 42.940(13)\\
       250000&   0.21160(7) & 61.18(2)\\
       400000&   0.34300(13) & 100.60(4)\\
       500000&   0.43010(13) & 126.70(4)\\
      1000000&   0.86550(19) & 256.50(6)\\
      2000000&   1.7350(4) & 515.40(11)\\
      4000000&   3.4740(8) & 1033.0(2)\\
 \hline
   \end{tabular}
   \label{tab1}
\end{table}
 
 \begin{table}
 \caption{TABLE I. ({\sl Continued.}) } 
  \begin{tabular}{cccc}
  \toprule
  Temperature  &  Pressure  &  Internal energy \\
  (K)  &  (Mbar)  &  (eV$/$atom) \\
 \hline 
\multicolumn{3}{c}{ $\rho$ = 2.49451$\times 10^{-2}$g$\,$cm$^{-3}$ [$r_s =6.0$~$a_0$]}\\
        15625&   0.01485(12) & -11.820(17)\\
        31250&   0.03556(11) & -6.948(13)\\
        62500&   0.09550(10) & 4.773(12)\\
        95250&   0.16610(10) & 15.740(13)\\
       125000&   0.23050(11) & 24.790(14)\\
       181825&   0.35170(12) & 40.920(16)\\
       250000&   0.4946(2) & 59.40(3)\\
       400000&   0.8072(4) & 99.19(5)\\
       500000&   1.0150(3) & 125.40(4)\\
      1000000&   2.0480(5) & 255.50(6)\\
      2000000&   4.1090(9) & 514.50(12)\\
      4000000&   8.2310(17) & 1032.0(2)\\
\multicolumn{3}{c}{ $\rho$ = 4.31052$\times 10^{-2}$g$\,$cm$^{-3}$ [$r_s =5.0$~$a_0$]}\\
        15625&   0.02537(20) & -11.970(16)\\
        31250&   0.05980(19) & -7.537(14)\\
        62500&   0.15800(16) & 3.087(12)\\
        95250&   0.27800(18) & 13.820(13)\\
       125000&   0.3880(3) & 22.910(18)\\
       181825&   0.5976(2) & 39.250(17)\\
       250000&   0.8460(3) & 58.01(3)\\
       400000&   1.3870(4) & 97.96(3)\\
       500000&   1.7460(5) & 124.40(4)\\
      1000000&   3.5310(8) & 254.50(6)\\
      2000000&   7.0960(13) & 513.80(10)\\
      4000000&   14.220(3) & 1031.0(2)\\
\multicolumn{3}{c}{ $\rho$ = 8.41898$\times 10^{-2}$g$\,$cm$^{-3}$ [$r_s =4.0$~$a_0$]}\\
        15625&   0.0479(7) & -12.21(2)\\
        31250&   0.1150(5) & -8.135(19)\\
        62500&   0.2950(6) & 1.17(2)\\
        95250&   0.5200(6) & 11.37(2)\\
       125000&   0.7308(7) & 20.29(3)\\
       181825&   1.1400(6) & 36.80(2)\\
       250000&   1.6250(9) & 55.71(3)\\
       400000&   2.6850(11) & 96.11(4)\\
       500000&   3.3910(11) & 122.70(4)\\
      1000000&   6.8810(15) & 253.30(6)\\
      2000000&   13.850(3) & 513.00(11)\\
      4000000&   27.750(6) & 1030.0(2)\\
\multicolumn{3}{c}{ $\rho$ = 0.1 g$\,$cm$^{-3}$ [$r_s \simeq 3.777$~$a_0$]}\\
        15625&   0.0578(15) & -12.21(5)\\
        31250&   0.1351(6) & -8.351(19)\\
        62500&   0.3491(6) & 0.74(2)\\
        95250&   0.6110(6) & 10.690(18)\\
       125000&   0.8598(8) & 19.57(2)\\
       181825&   1.3450(9) & 36.08(3)\\
       250000&   1.9200(9) & 55.00(3)\\
       400000&   3.1850(13) & 95.67(4)\\
       500000&   4.0180(11) & 122.10(3)\\
      1000000&   8.1680(17) & 252.90(5)\\
      2000000&   16.440(3) & 512.40(11)\\
      4000000&   32.970(7) & 1030.0(2)\\
 \hline
   \end{tabular}
  \end{table}
 
 \begin{table}
 \caption{TABLE I. ({\sl Continued.}) } 
  \begin{tabular}{cccc}
  \toprule
  Temperature  &  Pressure  &  Internal energy \\
  (K)  &  (Mbar)  &  (eV$/$atom) \\
 \hline 
\multicolumn{3}{c}{ $\rho$ = 0.199561 g$\,$cm$^{-3}$ [$r_s =3.0$~$a_0$]}\\
        15625&   0.124(3) & -12.31(4)\\
        31250&   0.2740(17) & -8.91(3)\\
        62500&   0.6730(17) & -1.11(3)\\
        95250&   1.1760(16) & 8.14(3)\\
       125000&   1.656(2) & 16.66(3)\\
       181825&   2.6060(15) & 32.91(2)\\
       250000&   3.753(2) & 51.98(3)\\
       400000&   6.273(3) & 92.85(4)\\
       500000&   7.943(2) & 119.60(4)\\
      1000000&   16.250(5) & 251.10(8)\\
      2000000&   32.760(7) & 510.90(11)\\
      4000000&   65.740(14) & 1029.0(2)\\
\multicolumn{3}{c}{ $\rho$ = 0.306563 g$\,$cm$^{-3}$ [$r_s =2.6$~$a_0$]}\\
        15625&   0.219(5) & -12.29(5)\\
        31250&   0.447(4) & -9.15(4)\\
        62500&   1.048(4) & -1.90(4)\\
        95250&   1.781(5) & 6.68(5)\\
       125000&   2.509(4) & 14.98(4)\\
       181825&   3.947(3) & 30.97(4)\\
       250000&   5.693(5) & 49.91(5)\\
       400000&   9.558(8) & 90.87(8)\\
       500000&   12.120(6) & 117.70(6)\\
      1000000&   24.890(9) & 249.60(9)\\
      2000000&   50.250(13) & 509.70(13)\\
      4000000&   101.100(19) & 1030.0(2)\\
\multicolumn{3}{c}{ $\rho$ = 0.389768 g$\,$cm$^{-3}$ [$r_s =2.4$~$a_0$]}\\
        15625&   0.298(12) & -12.30(10)\\
        31250&   0.597(9) & -9.21(7)\\
        62500&   1.337(8) & -2.40(7)\\
        95250&   2.280(7) & 6.11(6)\\
       125000&   3.175(8) & 14.09(7)\\
       181825&   4.979(11) & 29.84(9)\\
       250000&   7.206(9) & 48.84(7)\\
       400000&   12.090(12) & 89.61(9)\\
       500000&   15.370(14) & 116.70(11)\\
      1000000&   31.580(14) & 248.60(11)\\
      2000000&   63.96(2) & 509.80(19)\\
      4000000&   128.40(3) & 1028.0(3)\\
\multicolumn{3}{c}{ $\rho$ = 0.506024 g$\,$cm$^{-3}$ [$r_s =2.2$~$a_0$]}\\
        15625&   0.42(3) & -12.1(2)\\
        31250&   0.849(14) & -9.16(9)\\
        62500&   1.789(12) & -2.68(7)\\
        95250&   2.954(10) & 5.30(6)\\
       125000&   4.088(12) & 13.02(7)\\
       181825&   6.396(10) & 28.48(6)\\
       250000&   9.243(12) & 47.20(7)\\
       400000&   15.620(14) & 88.25(9)\\
       500000&   19.84(3) & 115.10(18)\\
      1000000&   40.94(3) & 247.5(2)\\
      2000000&   82.93(3) & 508.6(2)\\
      4000000&   166.50(4) & 1026.0(3)\\
 \hline
   \end{tabular}
  \end{table}
 
 \begin{table}
 \caption{TABLE I. ({\sl Continued.}) } 
  \begin{tabular}{ccc}
  \toprule
  Temperature  &  Pressure  &  Internal energy \\
  (K)  &  (Mbar)  &  (eV$/$atom) \\
 \hline 
\multicolumn{3}{c}{ $\rho$ = 0.673518 g$\,$cm$^{-3}$  [$r_s =2.0$~$a_0$]}\\
        15625&   0.59(4) & -12.02(18)\\
        31250&   1.28(2) & -8.96(11)\\
        62500&   2.461(10) & -3.01(5)\\
        95250&   3.930(7) & 4.43(3)\\
       125000&   5.413(6) & 11.91(3)\\
       181825&   8.446(8) & 27.08(4)\\
       250000&   12.200(7) & 45.63(3)\\
       400000&   20.660(14) & 86.64(6)\\
       500000&   26.270(15) & 113.50(7)\\
      1000000&   54.300(20) & 245.90(9)\\
      2000000&   110.20(3) & 507.20(16)\\
      4000000&   221.60(6) & 1026.0(3)\\
\multicolumn{3}{c}{ $\rho$ = 0.837338 g$\,$cm$^{-3}$ [$r_s =1.86$~$a_0$]}\\
        15625&   0.97(6) & -11.3(2)\\
        31250&   1.71(2) & -8.85(8)\\
        62500&   3.17(4) & -3.17(14)\\
        95250&   5.04(4) & 4.30(14)\\
       125000&   6.80(3) & 11.43(11)\\
       181825&   10.52(2) & 26.29(9)\\
       250000&   15.15(3) & 44.67(10)\\
       400000&   25.62(4) & 85.52(16)\\
       500000&   32.67(4) & 112.70(17)\\
      1000000&   67.44(6) & 244.9(2)\\
      2000000&   136.90(11) & 506.2(4)\\
      4000000&   275.50(8) & 1026.0(3)\\
\multicolumn{3}{c}{ $\rho$ = 1.0 g$\,$cm$^{-3}$  [$r_s =1.753$~$a_0$]}\\
        15625&   1.33(7) & -11.0(2)\\
        31250&   2.22(4) & -8.67(12)\\
        62500&   3.92(4) & -3.23(14)\\
        95250&   6.06(4) & 3.88(13)\\
       125000&   8.16(4) & 10.90(11)\\
       181825&   12.57(3) & 25.60(10)\\
       250000&   18.07(3) & 43.85(10)\\
       400000&   30.36(3) & 84.00(9)\\
       500000&   38.81(3) & 111.30(10)\\
      1000000&   80.40(4) & 243.90(13)\\
      2000000&   163.20(6) & 505.00(17)\\
      4000000&   328.80(8) & 1024.0(3)\\
\multicolumn{3}{c}{ $\rho$ = 1.00537 g$\,$cm$^{-3}$ [$r_s =1.750$~$a_0$]}\\
        15625&   1.35(9) & -10.9(3)\\
        31250&   2.23(3) & -8.69(10)\\
        62500&   4.03(4) & -2.97(14)\\
        95250&   6.03(5) & 3.67(16)\\
       125000&   8.27(5) & 11.09(14)\\
       181825&   12.63(3) & 25.56(9)\\
       250000&   18.17(4) & 43.82(13)\\
       400000&   30.54(5) & 84.05(16)\\
       500000&   38.99(8) & 111.2(3)\\
      1000000&   80.80(7) & 243.7(2)\\
      2000000&   164.20(11) & 505.3(3)\\
      4000000&   330.50(12) & 1024.0(4)\\
 \hline
   \end{tabular}
  \end{table}

 \begin{table}
 \caption{TABLE I. ({\sl Continued.}) } 
  \begin{tabular}{ccc}
  \toprule
  Temperature  &  Pressure  &  Internal energy \\
  (K)  &  (Mbar)  &  (eV$/$atom) \\
 \hline 
\multicolumn{3}{c}{ $\rho$ = 1.15688 g$\,$cm$^{-3}$ [$r_s =1.67$~$a_0$]}\\
        15625&   1.67(11) & -10.9(3)\\
        31250&   2.78(5) & -8.35(14)\\
        62500&   4.82(6) & -2.87(16)\\
        95250&   7.05(6) & 3.49(17)\\
       125000&   9.65(6) & 10.93(17)\\
       181825&   14.45(6) & 24.77(17)\\
       250000&   20.81(5) & 42.94(14)\\
       400000&   35.12(5) & 83.36(13)\\
       500000&   44.82(3) & 110.50(9)\\
      1000000&   92.99(5) & 243.30(13)\\
      2000000&   189.00(11) & 505.0(3)\\
      4000000&   380.30(16) & 1024.0(4)\\
\multicolumn{3}{c}{ $\rho$ = 1.31547 g$\,$cm$^{-3}$ [$r_s =1.60$~$a_0$]}\\
        31250&   3.46(9) & -7.9(2)\\
        62500&   5.65(10) & -2.8(2)\\
        95250&   8.31(8) & 3.75(19)\\
       125000&   10.97(6) & 10.44(15)\\
       181850&   16.66(6) & 24.76(14)\\
       250000&   23.73(5) & 42.50(12)\\
       400000&   39.92(7) & 82.72(17)\\
       500000&   50.66(7) & 109.20(17)\\
      1000000&   105.50(8) & 242.30(20)\\
      2000000&   214.20(7) & 502.90(16)\\
      4000000&   431.80(11) & 1022.0(3)\\
\multicolumn{3}{c}{ $\rho$ = 1.59649 g$\,$cm$^{-3}$ [$r_s =1.50$~$a_0$]}\\
        31250&   4.67(13) & -7.5(3)\\
        62500&   7.24(12) & -2.7(2)\\
        95250&   10.53(13) & 3.9(2)\\
       125000&   13.68(11) & 10.4(2)\\
       181850&   20.19(7) & 23.87(13)\\
       250000&   28.78(8) & 41.57(15)\\
       400000&   48.31(7) & 81.53(14)\\
       500000&   61.33(10) & 107.90(19)\\
      1000000&   128.20(11) & 241.8(2)\\
      2000000&   260.40(15) & 503.1(3)\\
      4000000&   524.30(13) & 1022.0(3)\\
\multicolumn{3}{c}{ $\rho$ = 1.96361 g$\,$cm$^{-3}$ [$r_s =1.40$~$a_0$]}\\
        31250&   6.4(2) & -7.0(4)\\
        62500&   9.69(15) & -2.1(2)\\
        95250&   13.66(14) & 4.3(2)\\
       125000&   17.11(10) & 10.05(16)\\
       181850&   25.29(6) & 23.68(9)\\
       250000&   35.66(9) & 41.00(14)\\
       400000&   59.23(9) & 80.18(15)\\
       500000&   75.41(10) & 106.90(15)\\
      1000000&   156.80(9) & 239.50(15)\\
      2000000&   319.49(16) & 501.2(3)\\
      4000000&   644.54(18) & 1021.0(3)\\
 \hline
   \end{tabular}
  \end{table}
 
 \begin{table}
 \caption{TABLE I. ({\sl Continued.}) } 
  \begin{tabular}{ccc}
  \toprule
  Temperature  &  Pressure  &  Internal energy \\
  (K)  &  (Mbar)  &  (eV$/$atom) \\
 \hline 
\multicolumn{3}{c}{ $\rho$ = 2.45250 g$\,$cm$^{-3}$ [$r_s =1.30$~$a_0$]}\\
        62500&   12.61(17) & -2.2(2)\\
        95250&   18.18(16) & 4.9(2)\\
       125000&   22.09(15) & 10.05(19)\\
       181850&   32.63(10) & 24.00(13)\\
       250000&   45.03(12) & 40.55(15)\\
       400000&   74.54(10) & 79.73(13)\\
       500000&   93.88(11) & 105.30(14)\\
      1000000&   195.60(17) & 238.1(2)\\
      2000000&   398.20(18) & 499.4(2)\\
      4000000&   804.2(3) & 1020.0(4)\\
\multicolumn{3}{c}{ $\rho$ = 3.11814 g$\,$cm$^{-3}$ [$r_s =1.20$~$a_0$]}\\
        62500&   18.0(4) & -1.4(4)\\
        95250&   24.6(3) & 5.4(4)\\
       125000&   30.1(5) & 11.0(5)\\
       181850&   43.0(3) & 24.4(4)\\
       250000&   57.8(4) & 39.9(4)\\
       400000&   94.9(2) & 78.5(2)\\
       500000&   120.10(14) & 104.70(14)\\
      1000000&   248.4(2) & 236.7(3)\\
      2000000&   504.5(6) & 496.7(6)\\
      4000000&   1021.0(5) & 1018.0(5)\\
\multicolumn{3}{c}{ $\rho$ = 4.04819 g$\,$cm$^{-3}$ [$r_s =1.10$~$a_0$]}\\
        62500&   26.2(1.1) & -0.1(8)\\
        95250&   34.4(5) & 6.1(4)\\
       125000&   41.9(6) & 12.1(5)\\
       181850&   58.8(6) & 25.3(5)\\
       250000&   75.8(4) & 39.0(3)\\
       400000&   125.1(3) & 78.5(2)\\
       500000&   156.8(2) & 103.80(16)\\
      1000000&   321.0(3) & 234.0(3)\\
      2000000&   651.5(7) & 492.9(5)\\
      4000000&   1327.0(8) & 1018.0(6)\\
\multicolumn{3}{c}{ $\rho$ = 5.38815 g$\,$cm$^{-3}$ [$r_s =1.00$~$a_0$]}\\
        95250&   51.9(1.6) & 8.3(9)\\
       125000&   61.1(8) & 13.7(5)\\
       181850&   81.8(1.2) & 25.9(7)\\
       250000&   105.4(8) & 40.0(5)\\
       400000&   169.1(1.3) & 78.2(7)\\
       500000&   212.2(1.3) & 104.1(8)\\
      1000000&   429.4(1.1) & 233.5(6)\\
      2000000&   867.4(1.2) & 491.5(7)\\
      4000000&   1768.0(1.0) & 1018.0(6)\\
\multicolumn{3}{c}{ $\rho$ = 7.39115 g$\,$cm$^{-3}$ [$r_s =0.90$~$a_0$]}\\
        95250&   80(2) & 10.3(9)\\
       125000&   92(2) & 15.5(9)\\
       181850&   120(2) & 27.4(1.0)\\
       250000&   153.4(1.3) & 41.9(6)\\
       400000&   236.6(1.5) & 78.2(6)\\
       500000&   297.2(1.2) & 104.5(5)\\
      1000000&   590.2(1.4) & 231.8(6)\\
      2000000&   1183.0(1.6) & 486.5(7)\\
      4000000&   2422.0(1.6) & 1015.0(7)\\
 \hline
   \end{tabular}
  \end{table}
 
 \begin{table}
 \caption{TABLE I. ({\sl Continued.}) } 
  \begin{tabular}{ccc}
  \toprule
  Temperature  &  Pressure  &  Internal energy \\
  (K)  &  (Mbar)  &  (eV$/$atom) \\
 \hline 
\multicolumn{3}{c}{ $\rho$ = 10.0000 g$\,$cm$^{-3}$ [$r_s \simeq 0.81373$~$a_0$]}\\
       125000&   142(3) & 19.0(8)\\
       181850&   182(2) & 31.8(7)\\
       250000&   225.7(1.7) & 44.5(5)\\
       400000&   334(4) & 80.5(1.3)\\
       500000&   414.4(1.6) & 106.2(5)\\
      1000000&   802(2) & 230.5(6)\\
      2000000&   1596.0(1.6) & 483.3(5)\\
      4000000&   3276.0(1.6) & 1013.0(5)\\
      8000000&   6592(3) & 2054.0(8)\\
\multicolumn{3}{c}{ $\rho$ = 10.5237 g$\,$cm$^{-3}$  [$r_s =0.80$~$a_0$]}\\
       125000&   153(7) & 19.8(2.0)\\
       181850&   197(4) & 33.1(1.0)\\
       250000&   242(3) & 45.4(9)\\
       400000&   353(4) & 80.4(1.2)\\
       500000&   434(4) & 105.3(1.1)\\
      1000000&   846(3) & 231.0(8)\\
      2000000&   1681(2) & 483.2(7)\\
      4000000&   3447(6) & 1011.0(1.8)\\
      8000000&   6929(3) & 2051.0(9)\\
\multicolumn{3}{c}{ $\rho$ = 15.7089 g$\,$cm$^{-3}$ [$r_s =0.70$~$a_0$]}\\
       181850&   346(9) & 40.0(1.7)\\
       250000&   419(9) & 55.0(1.8)\\
       400000&   575(4) & 87.1(8)\\
       500000&   684(3) & 109.3(6)\\
      1000000&   1293(4) & 233.5(7)\\
      2000000&   2515(4) & 481.4(9)\\
      4000000&   5149(5) & 1011.0(1.1)\\
      8000000&   10390(5) & 2060.0(1.1)\\
\multicolumn{3}{c}{ $\rho$ = 24.9451 g$\,$cm$^{-3}$ [$r_s =0.60$~$a_0$]}\\
       400000&   1037(12) & 98.4(1.4)\\
       500000&   1208(17) & 121(2)\\
      1000000&   2133(11) & 239.2(1.4)\\
      2000000&   4025(9) & 480.9(1.1)\\
      4000000&   8195(16) & 1009.0(2.0)\\
      8000000&   16200(17) & 2020(2)\\
     16000000&   32950(13) & 4124.0(1.6)\\
\multicolumn{3}{c}{ $\rho$ = 43.1052 g$\,$cm$^{-3}$ [$r_s =0.50$~$a_0$]}\\
       400000&   2212(30) & 123(2)\\
       500000&   2523(20) & 146.5(1.8)\\
      1000000&   4002(30) & 256.4(2.0)\\
      2000000&   7162(18) & 490.2(1.3)\\
      4000000&   14180(17) & 1006.0(1.3)\\
      8000000&   28390(30) & 2044(2)\\
     16000000&   56880(20) & 4118.0(1.7)\\
     32000000&   114000(40) & 8273(3)\\
     64000000&   227900(90) & 16540(7)\\
\multicolumn{3}{c}{ $\rho$ = 84.1898 g$\,$cm$^{-3}$ [$r_s =0.40$~$a_0$]}\\
      1000000&   9169(50) & 298.4(2.0)\\
      2000000&   14950(80) & 517(3)\\
      4000000&   27960(70) & 1007(3)\\
      8000000&   54980(40) & 2019.0(1.4)\\
     16000000&   110600(70) & 4093(3)\\
     32000000&   222600(120) & 8262(4)\\
     64000000&   445900(110) & 16570(4)\\
 \hline
   \end{tabular}
  \end{table}
 
 \begin{table}
 \caption{TABLE I. ({\sl Continued.}) } 
  \begin{tabular}{cccc}
  \toprule
  Temperature  &  Pressure  &  Internal-energy \\
  (K)  &  (Mbar)  &  (eV$/$atom) \\
 \hline 
\multicolumn{3}{c}{ $\rho$ = 199.561 g$\,$cm$^{-3}$ [$r_s =0.30$~$a_0$]}\\
      2000000&   41350(800) & 597(12)\\
      4000000&   70230(400) & 1056(6)\\
      8000000&   129900(500) & 2000(8)\\
     16000000&   263400(300) & 4104(5)\\
     32000000&   527000(400) & 8247(7)\\
     64000000&   1049000(400) & 16450(7)\\
\multicolumn{3}{c}{ $\rho$ = 673.518 g$\,$cm$^{-3}$ [$r_s =0.20$~$a_0$]}\\
      4000000&   299900(4000) & 1322(16)\\
      8000000&   504200(3000) & 2281(15)\\
     16000000&   897500(1300) & 4121(6)\\
     32000000&   1783000(1700) & 8251(8)\\
     64000000&   3569000(1600) & 16560(7)\\
\multicolumn{3}{c}{ $\rho$ = 1596.49 g$\,$cm$^{-3}$ [$r_s =0.15$~$a_0$]}\\
      4000000&   1071000(20000) & 2002(40)\\
      8000000&   1555000(20000) & 2961(40)\\
     16000000&   2342000(17000) & 4517(30)\\
     32000000&   4565000(16000) & 8890(30)\\
     64000000&   8523000(8000) & 16670(17)\\
 \botrule
   \end{tabular}
  \end{table}

\section{Particle Correlations}

\begin{figure}
\rotatebox{0}{\includegraphics[scale=1.0]{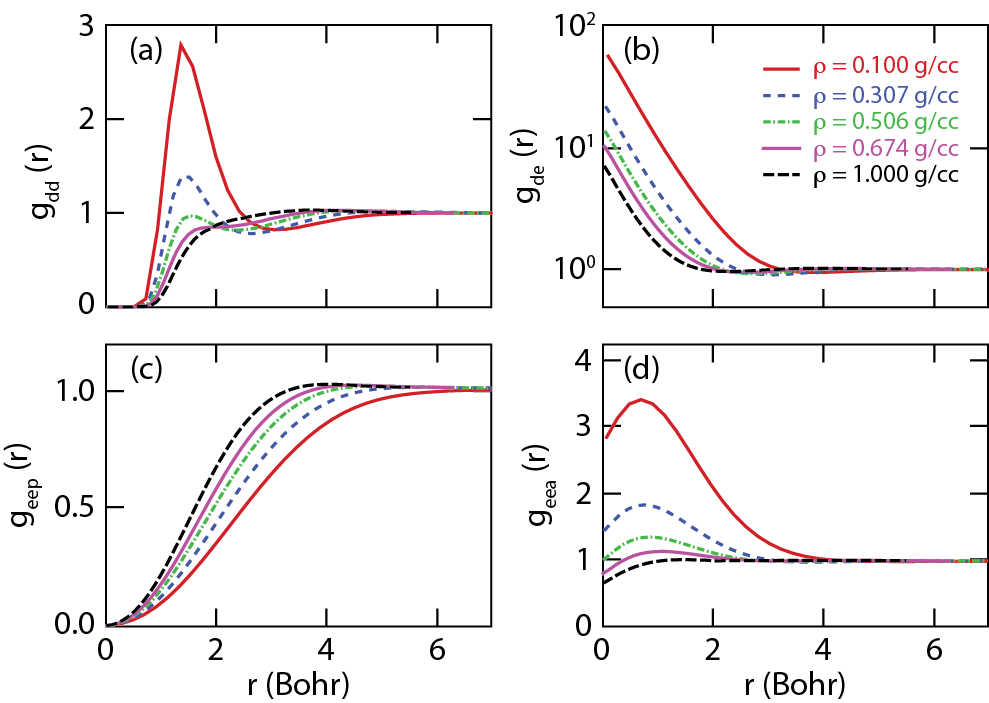}}
\caption{(Color online) The pair correlation functions $g(r)$ derived from PIMC calculations: 
(a) the ion-ion correlation $g_{dd}(r)$; 
(b) the ion-electron correlation $g_{de}(r)$;
(c) the electron-electron correlation $g_{eep}(r)$ for parallel spins; 
(d) the electron-electron correlation $g_{eea}(r)$ for anti-parallel spins, 
with different densities at 15$\,$625 K.
}
\label{gofr_T_1.35eV}
\end{figure}

\begin{figure}
\rotatebox{0}{\includegraphics[scale=1.0]{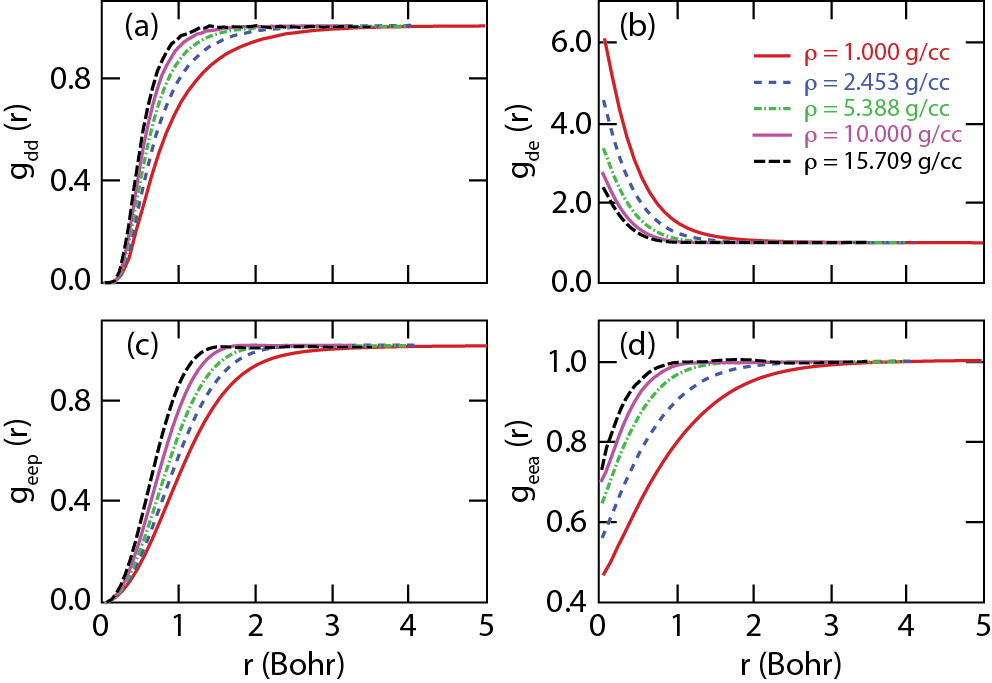}}
\caption{(Color online) Pair correlation functions similar to Fig.~\ref{gofr_T_1.35eV} 
but at a higher temperature of $ 2.5 \times 10^5$ K and densities from 1.0 to 15.709 \gcc.}
\label{gofr_T_21.54eV}
\end{figure}

\begin{figure}
\rotatebox{0}{\includegraphics[scale=1.0]{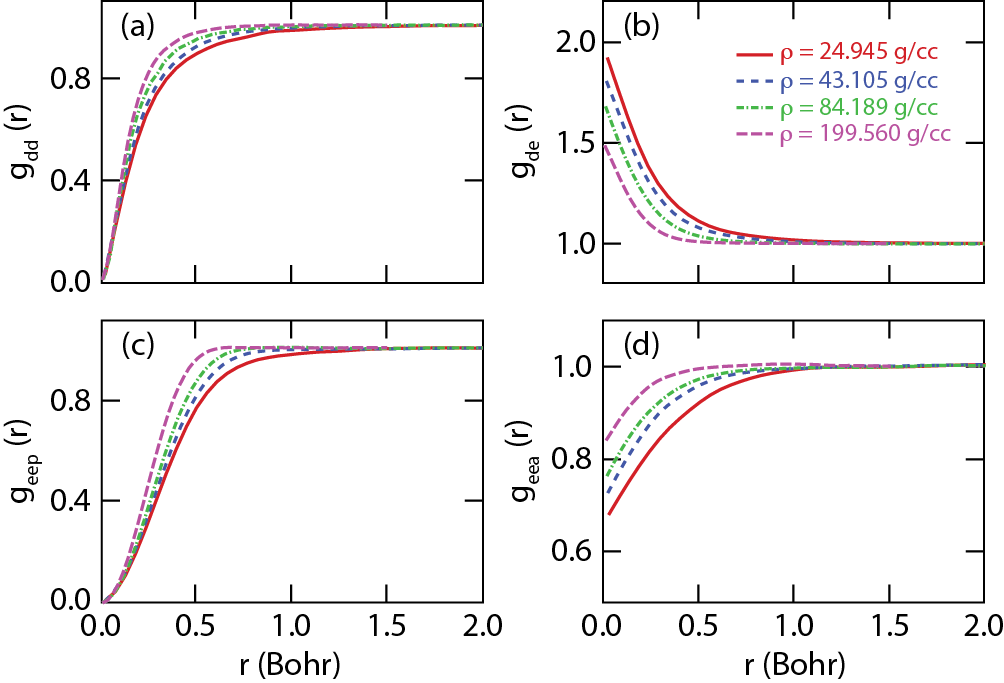}}
\caption{(Color online) Pair correlation functions similar to Fig.~\ref{gofr_T_1.35eV} 
but at a higher temperature of $2\times 10^6$~K and densities from 24.945 to 199.56 \gcc.}
\label{gofr_T_172.35eV}
\end{figure}

The correlation functions, $g(r)$, between different pairs of
particles such as electron-electron, electron-ion, and ion-ion are
particularly interesting for analyzing the physical and chemical
changes in the plasma at various density and temperature conditions.
The $g(r)$ are available directly in PIMC simulations. We first show
the density effects on the structure of the fluid structure by showing
how the $g(r)$ functions change with density for three temperatures of
15$\,$625 K, $2.5\times 10^5$ K, and $2 \times 10^6$ K in
Figs.~\ref{gofr_T_1.35eV}-\ref{gofr_T_172.35eV}. 

Fig.~\ref{gofr_T_1.35eV}(a) shows a clear peak in the ion-ion
correlation function, $g_{dd}(r)$, for 0.1 {\gcc} at the molecular bond
length of 1.4 $a_0$. As the density of deuterium increases to 1.0 \gcc, one
observes a drastic reduction in peak height which demonstrates the
pressure-induced dissociation of D$_2$ molecules, confirming earlier
PIMC results~\cite{Mi99,MiltzerPhD}. This interpretation is also
supported by the reduction of peak at $r=0$ in the $g_{de}(r)$
function in Fig.~\ref{gofr_T_1.35eV}(b). Furthermore the positive
correlation between pair of electrons with anti-parallel spin in
Fig.~\ref{gofr_T_1.35eV}(d) is also disappearing with increasing density
since they are no longer bound into molecules.
Fig.~\ref{gofr_T_1.35eV}(c) shows that there is always a strong repulsion
between electrons with parallel spins because of the Pauli exclusion
principle but they approach each other more at higher densities.

Figs.~\ref{gofr_T_21.54eV} and ~\ref{gofr_T_172.35eV} show the pair
correlation functions for different densities at much higher
temperatures of $2.5\times 10^5$ K and $2\times 10^6$ K.  At these
temperatures, $D_2$ molecules have completely dissociated as indicated
by the absence of the peak in the ion-ion correlation function.  The
attractive forces between pair of ions have disappeared and repulsion
now dominates their interactions. At higher densities, particles are
``packed'' more tightly and approach each other significantly more so
that the $g(r)$ rise up more steeply and reach the values of 0.5 as
much smaller distances.

\begin{figure}
\rotatebox{0}{\includegraphics[scale=1.0]{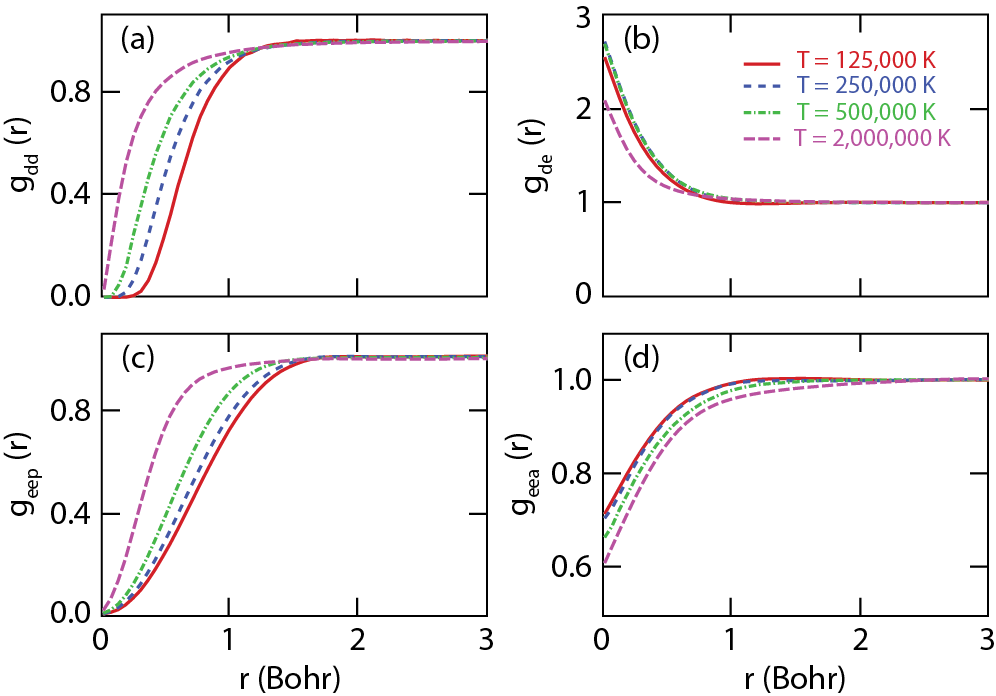}}
\caption{(Color online) Pair correlation functions similar to 
Fig.~\ref{gofr_T_1.35eV} at a fixed density of 10.0 {\gcc} for 
temperatures ranging from $1.25\times 10^5$~K to $2\times 10^6$~K.}
\label{gofr_rho_10.0gcc}
\end{figure}

In Fig.~\ref{gofr_rho_10.0gcc}, we compare the pair correlation
functions for the fixed density of 10 {\gcc} for temperatures ranging
from $1.25\times 10^5$ K to $2\times 10^6$ K. It is interesting to
note there is relatively little variation between the three curves
below the Fermi temperature of $T_F = 8.8\times 10^5$ K but they
differ significantly at $2\times 10^6$ K. This is a
manifestation of Fermi degeneracy effects in which the electrons
approach the ground state for temperatures well below the Fermi
temperature. Then much of the temperature dependence of the pair
correlation functions disappears. For example the pair correlation
functions of electrons with anti-parallel spins are almost identical
for the two lowest temperatures of $1.25\times 10^5$~K and $2.5\times
10^5$~K but they differ substantially from results at well above
$T_F$. When the temperature raises above $T_F$, Pauli exclusion
effects are reduced, the electrons start to occupy a variety of
states, which then has a positive feedback on the mobility of the
ions.

\section{Comparisons of the FPEOS table with {\sl SESAME} and {\sl Kerley03} models }

In this section, we compare the pressures and internal energies in our
{\sl FPEOS} table with predictions from the well-known
semi-analytical {\sl SESAME} and {\sl Kerley03} EOS tables. To
illustrate how much the system deviates from an ideal plasma, we have
normalized both pressure and energy to their corresponding 
values [$E_{id}$ and $P_{id}$] of
non-interacting gas of classical ions and fermionic electrons. This
removes most of the temperature dependence and emphasizes the effects
of the Coulomb interaction, which leads to a reduction in pressure and
energy below the non-interacting values in all cases.

\begin{figure}
\rotatebox{0}{\includegraphics[scale=1.0]{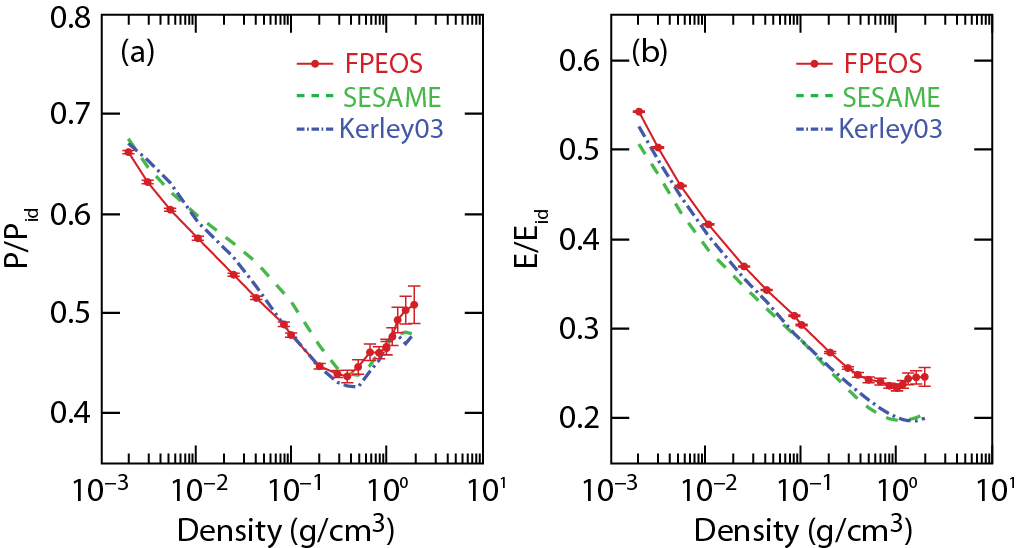}}
  \caption{(Color online) The comparisons of pressure (a) and internal
    energy (b) as a function of density from the the FPEOS, {\sl
      SESAME}, {\sl Kerley03} tables. The error bars indicate the
    1$\sigma$ statistical uncertainty in the PIMC simulations. Results
    were normalized to non-interacting gas of classical ions and
    fermionic electrons.}
\label{PE_comparison_2.69eV}
\end{figure}

\begin{figure}
\rotatebox{0}{\includegraphics[scale=1.0]{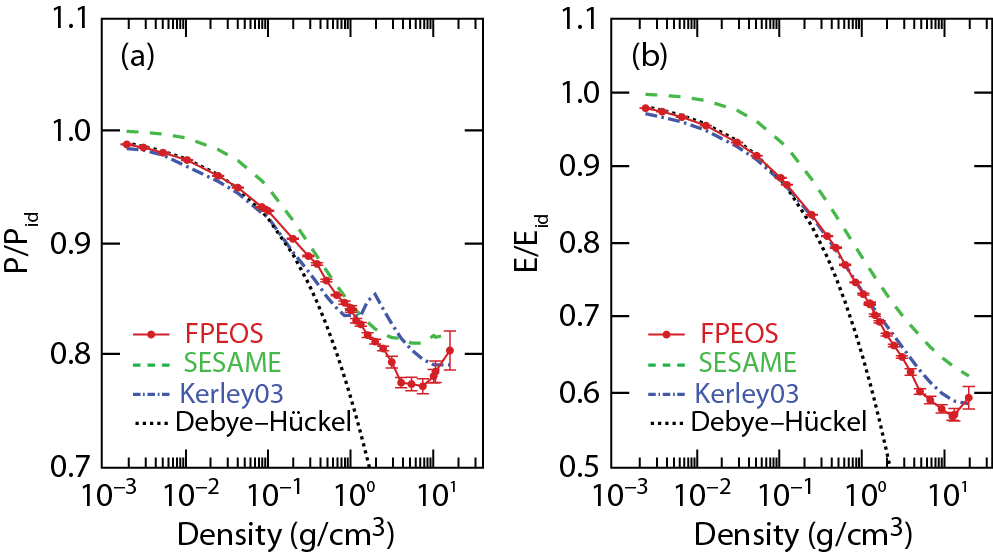}}
  \caption{(Color online) Same as Fig.~\ref{PE_comparison_2.69eV} but
    for a different temperature $2.5\times 10^5$~K.  The Debye-H\"uckel
    model is also shown for comparison.  }
\label{PE_comparison_21.54eV}
\end{figure}

\begin{figure}
\rotatebox{0}{\includegraphics[scale=1.0]{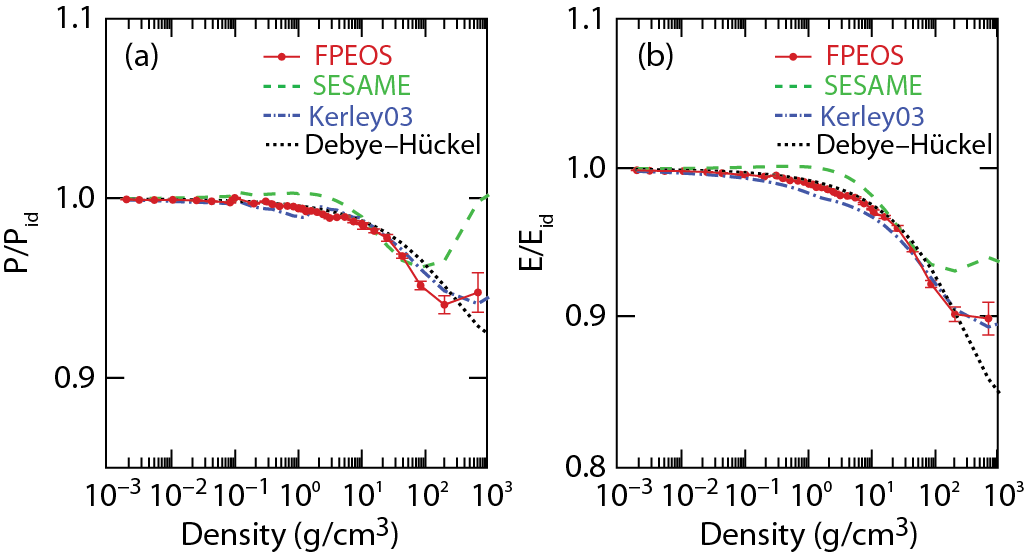}}
\caption{(Color online) Same as Fig.~\ref{PE_comparison_2.69eV} but for a 
different temperature of $4\times 10^6$~K.
}
\label{PE_comparison_344.69eV}
\end{figure}

\begin{figure}
\rotatebox{0}{\includegraphics[scale=1.0]{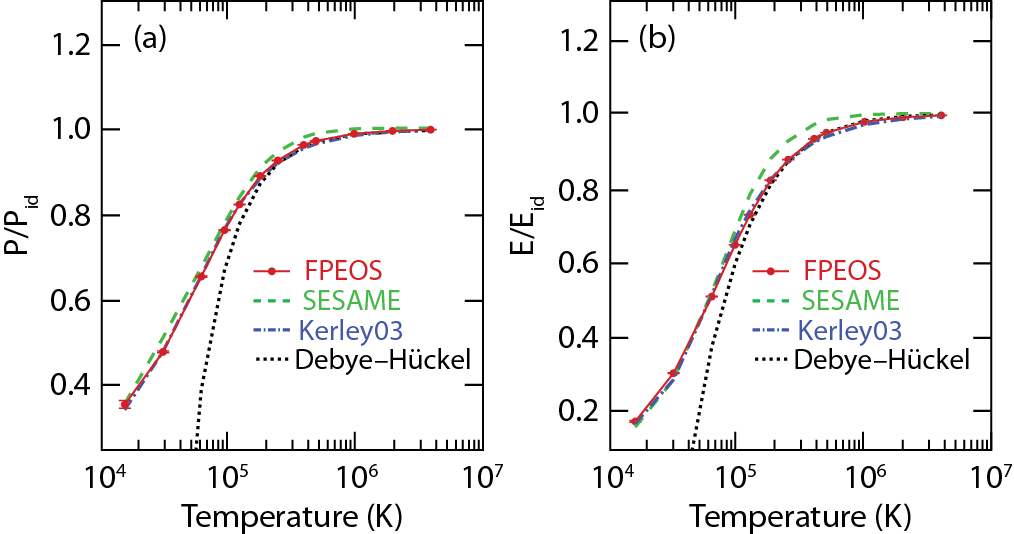}}
\caption{(Color online) Pressure (a) and energy (b) as a function of 
temperature from FPEOS, {\sl SESAME}, and {\sl Kerley03} tables for 
deuterium density of 0.1 \gcc.
}
\label{PE_comparison_0.1gcc}
\end{figure}

\begin{figure}
\rotatebox{0}{\includegraphics[scale=1.0]{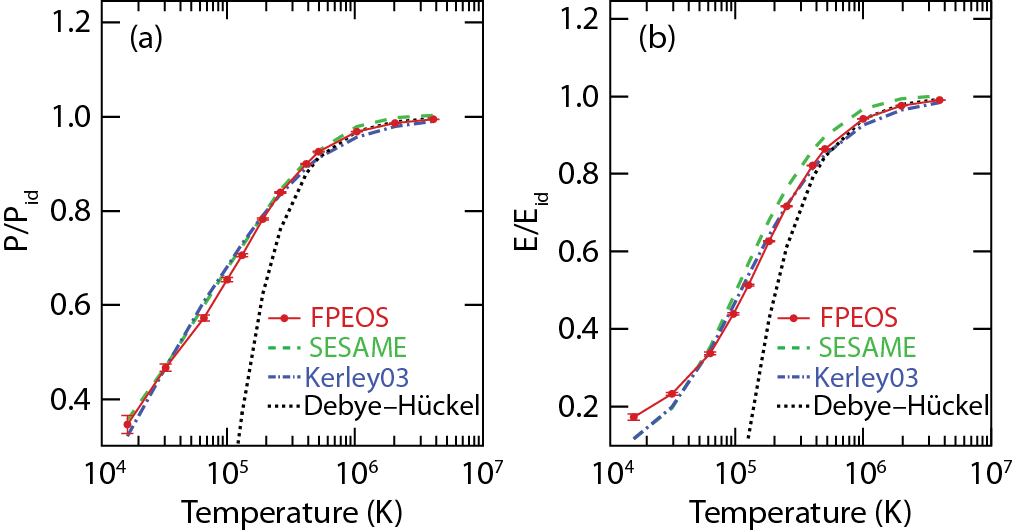}}
\caption{(Color online) Same as Fig.~\ref{PE_comparison_0.1gcc} but 
for a different deuterium density of 1.0 \gcc.
}
\label{PE_comparison_1.0gcc}
\end{figure}

\begin{figure}
\rotatebox{0}{\includegraphics[scale=1.0]{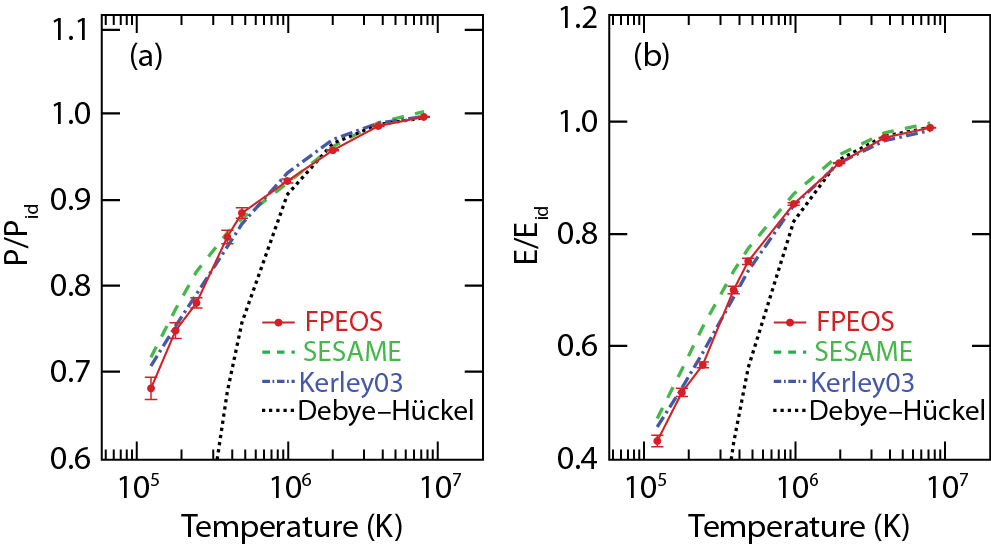}}
\caption{(Color online) Same as Fig.~\ref{PE_comparison_0.1gcc} but for 
a higher deuterium density of 10.0 \gcc.
}
\label{PE_comparison_10.0gcc}
\end{figure}

\begin{figure}
\rotatebox{0}{\includegraphics[scale=1.0]{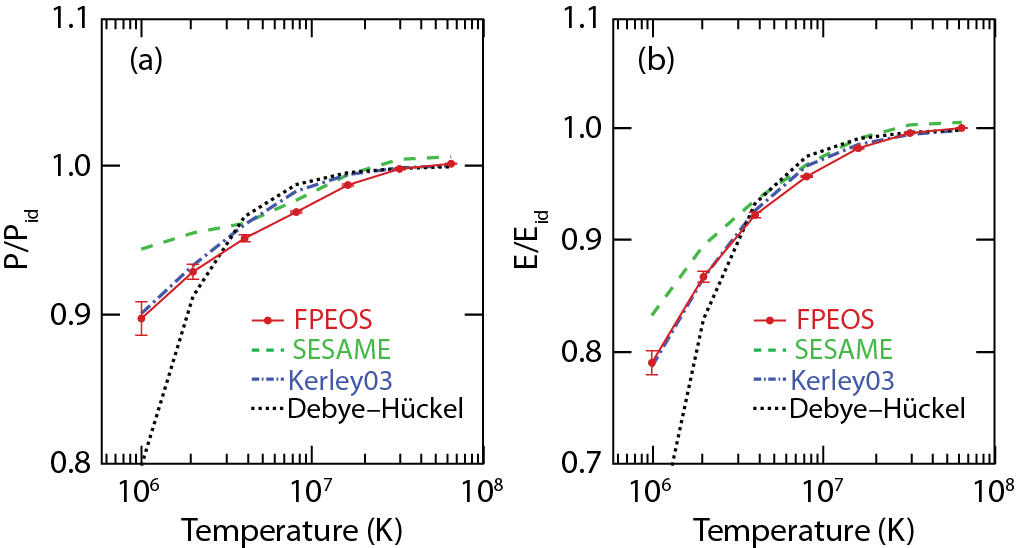}}
\caption{(Color online) Same as Fig.~\ref{PE_comparison_0.1gcc} but 
for much higher deuterium density 84.19 \gcc.
}
\label{PE_comparison_84.19gcc}
\end{figure}

In Figs.~\ref{PE_comparison_2.69eV}$-$\ref{PE_comparison_344.69eV}, we
plot the pressure and the internal energy as a function of density for
different temperatures ranging from $31250$ to $4\times 10^6$~K.
Figs.~\ref{PE_comparison_0.1gcc}$-$\ref{PE_comparison_84.19gcc} show
them as function of temperature for different densities varying
between 0.1 and 84.19 \gcc.

In Fig.~\ref{PE_comparison_2.69eV}, we compare FPEOS, {\sl SESAME},
and {\sl Kerley03} results at a comparatively low temperature of
31250~K. This is difficult regime to describe by chemical models
because the plasma consists of neutral species like molecules and
atoms as well as charged particles such as ions and free electrons.
The interaction between neutral and charged species is very difficult
analytically while it poses no major challenge to first-principles
simulations. As is shown by Fig.~\ref{PE_comparison_2.69eV}(a), the
{\sl SESAME} EOS predicts overall higher pressures at low density
($\rho \le 0.3$ \gcc) but then all three models come to agree with
each other at higher densities. The improved {\sl Kerley03} table
still showed some discrepancy at very low densities, even though some
improvements to the ionization equilibrium model have been
made~\cite{Kerley2003}.  

Fig.~\ref{PE_comparison_2.69eV}(b) shows that the internal energies
predicted by {\sl SESAME} and {\sl Kerley03} are overall lower than
the FPEOS values.  The higher the density, the more discrepancy there
is. Again, this manifests the difficulty of chemical models at such
plasma conditions.

One expects the pressure and internal energy to approach the values of
a non-interacting gas in the low-density and the high-density limit.
At low density, particles are so far away from each other that the
interaction effects become negligible. At high density, Pauli
exclusion effects dominate over all other interactions and all
thermodynamic function can be obtained from the ideal Fermi gas. Just
at an intermediate density range which still spans several orders of
magnitude, the Coulomb interaction matters and significant deviations
for the ideal behavior are observed.

For a higher temperature of $2.5\times 10^5$~K, the pressure and energy are 
compared in Fig.~\ref{PE_comparison_21.54eV}. The low-density deuterium at this 
temperature becomes fully ionized and can therefore be described by the Debye-H\"uckel 
plasma model \cite{DH-model}, which is based on the self-consistent 
solution of the Poisson equation for a system of screened charges. The pressure and 
energy per particle (counting electrons and ions) can be explicitly expressed as:
\begin{equation}
P_D = P_{id} - \frac{k_bT}{24\pi \lambda_D^3} \;\;{\rm and}\;\;
E_D = E_{id} - \frac{k_bT}{8\pi n \lambda_D^3}~,
\label{DH}
\end{equation}
with the particle number density $n$, the Boltzmann constant $k_b$, and
the Debye length $\lambda_D = \sqrt {k_bT/4\pi n e^2}$.  

We have added the Debye-H\"uckel results to
Figs.~\ref{PE_comparison_21.54eV}-\ref{PE_comparison_84.19gcc}. In
Fig.~\ref{PE_comparison_21.54eV} one finds that the simple
Debye-H\"uckel model perfectly agrees with our PIMC calculations in
the lower densities up to 0.1 \gcc, where the improved {\sl Kerley03}
EOS also gives very similar pressures and energies.  On the other
hand, the {\sl SESAME} EOS overestimates both pressure and energy even
at such low densities.

Fig.~\ref{PE_comparison_21.54eV}(a) exposes an artificial cusp in
pressure in {\sl Kerley03} EOS at densities of $1.5-4$ {\gcc} while the
internal energy curve is smooth. This artificial pressure cusp appears
for all temperatures at roughly the same density and may be related to
the artificial double compression peaks in the principal Hugoniot
predicted by {\sl Kerley03} EOS~\cite{Kerley2003}.
The Debye-H\"uckel model fails at densities higher than 0.2 {\gcc} for
this temperature. It is only applicable to weakly interacting plasmas
but otherwise predicts unphysically low pressures and energies.

As the temperature increased to $4\times 10^6$~K, the Debye-H\"uckel
model agrees very well with FPEOS in both pressure and energy over a
wide range of densities up to 20 {\gcc} as shown in
Fig.~\ref{PE_comparison_344.69eV}. Significant differences in both
pressure and energy are again found for the {\sl SESAME} EOS, when compared to FPEOS
and {\sl Kerley03} tables.  It should also be noted
that the internal energy predicted by {\sl Kerley03} is slightly lower
than those of FPEOS and the Debye-H\"uckel model for $\rho = 0.1-20$ \gcc.

In Figs.~\ref{PE_comparison_0.1gcc}$-$\ref{PE_comparison_84.19gcc}, we compare 
the pressure and energy versus temperature for specific densities 
of 0.1, 1.0, 10.0, 84.19 \gcc. At high temperature where the plasma is 
fully ionized, the Debye-H\"uckel model
well reproduces the FPEOS pressures and energies very well. It is interesting to note
that the {\sl SESAME} table overestimates the pressure and energy even for a 
fully ionized plasma at densities greater than 1.0 {\gcc} as shown in 
Figs.~\ref{PE_comparison_1.0gcc}$-$\ref{PE_comparison_84.19gcc}.   
For a very low density of 0.1 \gcc, Fig.~\ref{PE_comparison_0.1gcc}
shows that the improved {\sl Kerley03} agrees very well with FPEOS, while the 
{\sl SESAME} results are noticeably higher.
Moreover, the improvements made to {\sl Kerley03} have resulted in remarkable agreement 
with FPEOS for intermediate densities of 0.1 and 10.0 {\gcc}
depicted by Figs.~\ref{PE_comparison_1.0gcc} and \ref{PE_comparison_10.0gcc}.
Only a small deviation in the internal energy between {\sl Kerley03} and our FPEOS results 
can be found at the lowest temperature for 1.0 \gcc. 

At a higher density of 84.19 \gcc, the {\sl SESAME} EOS again
significantly deviates from both the FPEOS and the {\sl Kerley03} EOS
as is illustrated by Fig.~\ref{PE_comparison_84.19gcc}.  The latter
two EOS tables give very similar results in internal energy almost for
the entire temperature range, though the pressures predicted by {\sl
  Kerley03} are higher than the FPEOS ones for temperatures varying
from $2\times 10^6$~K to $2\times 10^7$~K.  In contrast to the
significant EOS differences seen from {\sl SESAME}, the improved {\sl
  Kerley03} table is overall in better agreement with the FPEOS table,
although subtle discrepancies and an artificial pressure cusp still
exist in the {\sl Kerley03} EOS.

\section{Applications to ICF}

With the EOS comparisons discussed above, we now investigate what differences can be observed
when these EOS tables are applied to simulate ICF shock timing experiments and target 
implosions. Using radiation-hydrodynamics codes, both one-dimensional {\sl LILAC}~\cite{LILAC} 
and two-dimensional {\sl DRACO}~\cite{draco}, for simulations of experiments, we can explore the 
implications of our first-principles equation of state table for the understanding and design of 
ICF targets.

\begin{figure}
\rotatebox{0}{\includegraphics[scale=1.0]{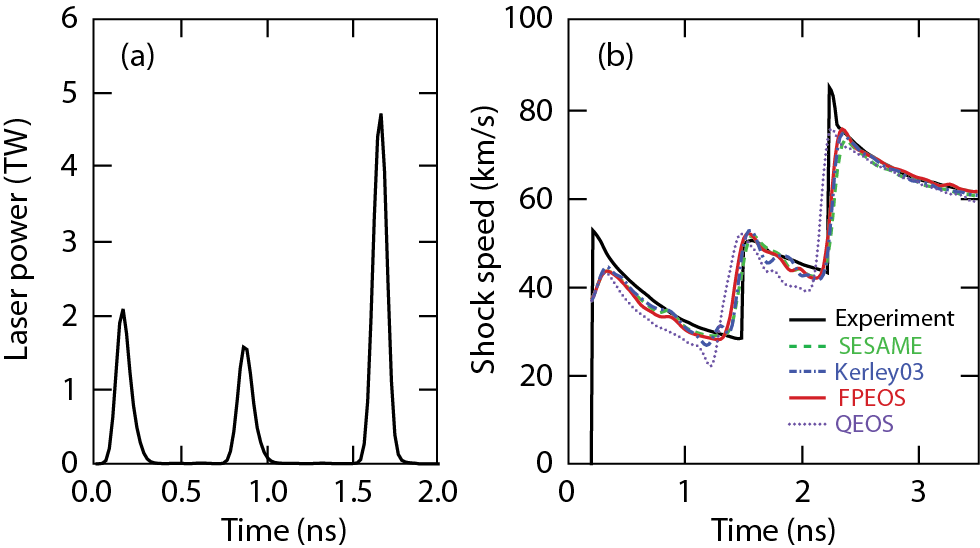}}
\caption{(Color online) (a) The triple-picket pulse shape for shock-timing experiments using 
cryogenic deuterium; and (b) the measured shock speed in liquid-D$_2$ (black solid-line) comparing 
with hydro-dynamics predictions using the {\sl SESAME}, FPEOS, {\sl Kerley03}, and {\sl QEOS} models.
}
\label{shock_timing}
\end{figure}

We first study the shock timing experiments
\cite{shock-timing1,shock-timing2} performed on the Omega laser
facility. As the fuel entropy in ICF implosions is set by a sequence
of shocks, the timing of shock waves in liquid deuterium is extremely
important for the ICF target performance. In shock timing experiments, the
carbon deuterium (CD) spherical shell, 900 $\mu m$ in diameter and 10
$\mu m$ thick, in a cone-in-shell geometry \cite{shock-timing1} were
filled with liquid deuterium. VISAR (velocity interferometery system
for any reflector) was used to measure the shock velocity. As is shown
in Fig.~\ref{shock_timing}(a), the triple-picket laser pulses are
designed to launch three shocks into the liquid deuterium. The
experimental results are plotted in Fig.~\ref{shock_timing}(b), in
which the shock front velocity is shown as a function of time.  One
finds that when the second shock catches up the first one at around
1.5 ns, the shock-front velocity exhibits a sudden jump.
Another velocity jump at $2.2$ ns occurs when the third strong
shock overtakes the previous two.  With the hydro-code {\sl LILAC}, we
have simulated the shock timing experiments using different EOS tables
including FPEOS, {\sl SESAME}, {\sl Kerley03}, and {\sl
  QEOS}~\cite{qeos}. The radiation hydro-simulations have used the
standard flux-limited ($f=0.06$) thermal transport model, although a
nonlocal model has resulted in better agreement with experiment for
the speed of first shock~\cite{shock-timing2}.  The results of FPEOS,
{\sl SESAME} and {\sl Kerley03} are in good agreement with the
experimental observation, while the {\sl QEOS} predicts s much lower
shock velocity and early catching up time. The shock timing
experiments can only explore a small range of deuterium densities
(0.6$-$2.5 \gcc) and temperatures ($3 - 10$ eV). In these plasma
conditions the {\sl SESAME} and {\sl Kerley03} have been
adjusted~\cite{Kerley2003} to match to the first-principles
calculations, which can be seen in Fig.~\ref{PE_comparison_1.0gcc}.
Thus, the shock velocity differences predicted by the FPEOS, {\sl
  SESAME}, and {\sl Kerley03} are very small in such plasma
conditions.

\begin{figure}
\rotatebox{0}{\includegraphics[scale=1.0]{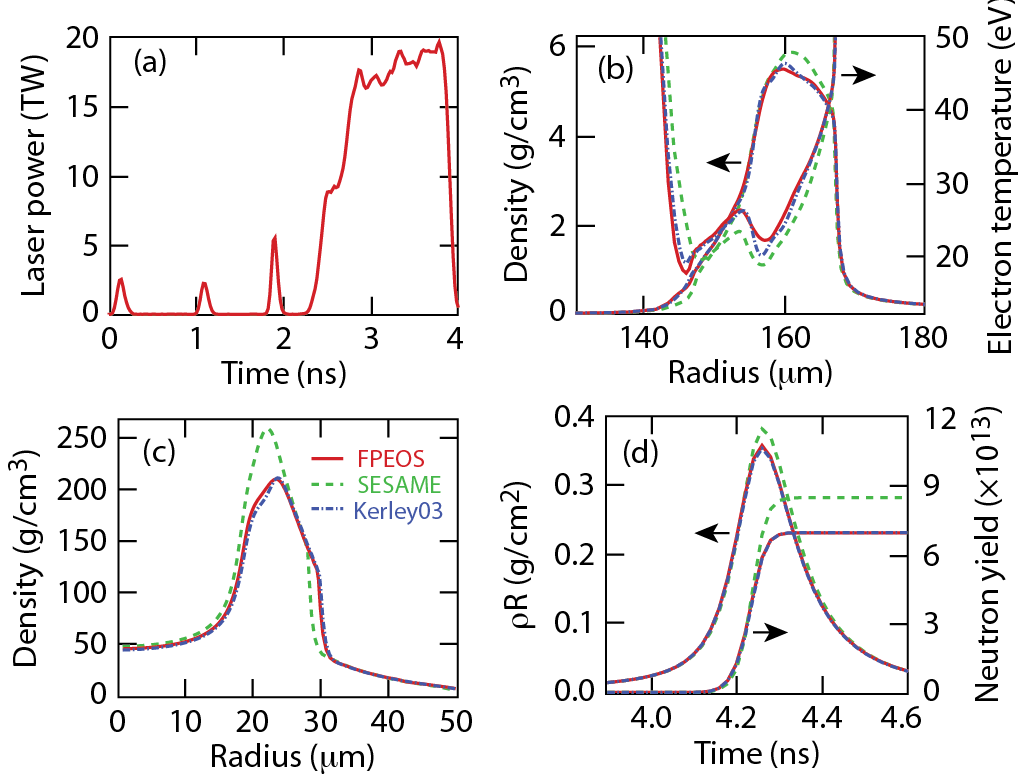}}
\caption{(Color online) The hydro-code simulations of a cryogenic DT implosion
on OMEGA using the three different EOS tables including {\sl SESAME}, FPEOS,
and {\sl Kerley03}: (a) The laser pulse shape; 
(b) The density-temperature profiles of the imploding DT shell at the 
middle of main laser pulse ($t=3.8$ ns);
(c) the density profile at the peak compression (t=4.26 ns); and (d) the 
areal density $\rho R$ and yield as a function of time.
}
\label{omega}
\end{figure}

\begin{figure}
\rotatebox{0}{\includegraphics[scale=1.0]{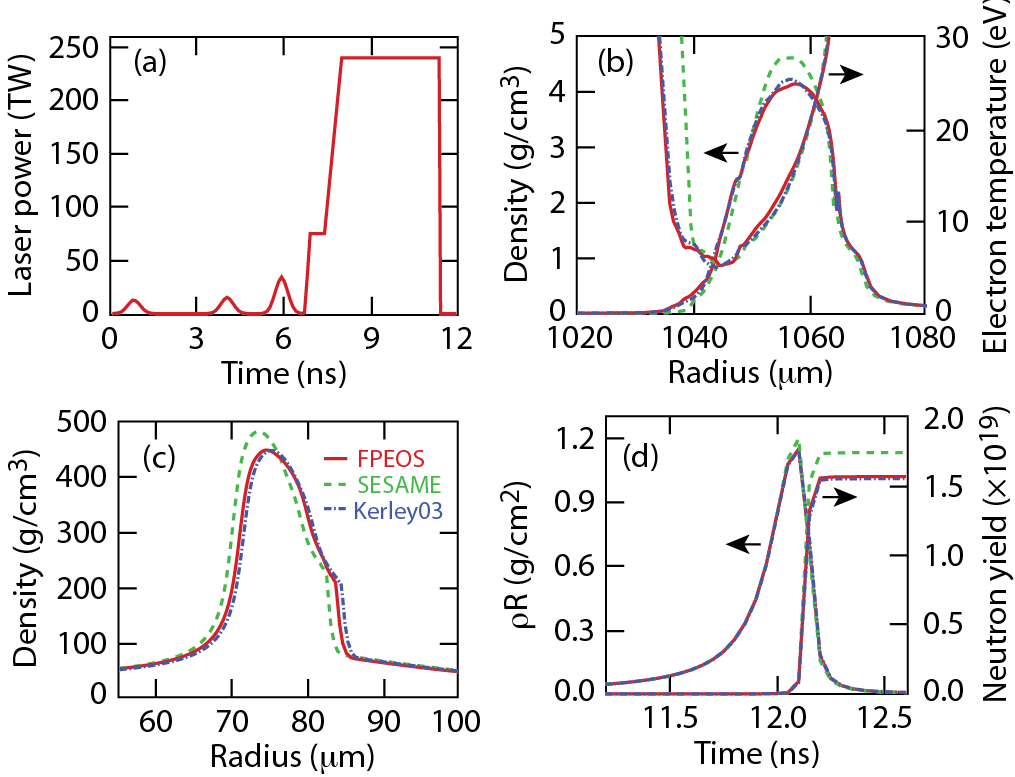}}
\caption{(Color online) Similar to Fig.~\ref{omega} but for a hydro-equivalent direct drive, 1 MJ 
ignition design for NIF. 
}
\label{nif}
\end{figure}

Next, we examine the implications of coupling and degeneracy effects in ICF implosions.
The possible differences in target compression and fusion yields 
of ICF implosions are investigated through radiation hydro-simulations using FPEOS 
in comparison to results predicted by {\sl SESAME} and {\sl Kerley03}.
The {\sl LILAC}-simulation results are compared in Figs.~\ref{omega} and \ref{nif}, 
respectively, for a $DT$ implosion on OMEGA and a hydro-equivalent direct-drive 
design on the NIF. In Figs.~\ref{omega}(a) and \ref{nif}(a), we plot the  
laser pulse shapes consisting of triple-pickets and the step-main-pulse. 
The cryogenic OMEGA DT target (860 $\mu m$ diameter) has a 10-$\mu m$ deuterated plastic
ablator and $\sim 65 ~\mu m$ of DT  ice. Figure ~\ref{omega}(b) shows the density
and temperature profiles at the end of the laser pulse ($t =3.8$ ns) from the FPEOS (red/solid line), 
the {\sl SESAME} (green/dashed line), and the {\sl Kerley03} (blue/dot-dashed line) simulations.
AT this time the shell has converged to a radius of $\sim 160~\mu m$ from its original 
radius of $\sim 430~\mu m$.
The shell?s peak density and average temperature
were $\rho \simeq 5.6$ g/cm$^3$ and $T \simeq 21$ eV, which correspond
to the coupled and degenerate regimes with $\Gamma \simeq 1.22$  and $\theta \simeq 0.47$. 
It is shown that the FPEOS simulation
predicted $\sim 10$\% lower peak density but $\sim 15$\% higher temperature
relative to the SESAME prediction. As is shown by the comparisons made 
in Fig. ~\ref{PE_comparison_21.54eV} and in Ref. [\cite{FPEOS_PRL}], 
 the FPEOS predicts slightly stiffer deuterium 
than {\sl SESAME} at the similar temperature regime.
 This explains the lower peak density seen in Fig.~\ref{omega}(b).
 The $\sim 15$\%  higher temperature in the
FPEOS case was originated from the lower internal energy
[see Fig. ~\ref{PE_comparison_21.54eV}(b)]. Since the laser ablation does the same
work/energy to the shell compression and its kinetic motion,
a lower internal energy in FPEOS means more energy
is partitioned to heat the shell, thereby resulting in a higher
temperature. Such a temperature increase and density drop
can have consequences in the implosion performance. 
Despite the subtle EOS differences discussed above, the {\sl Kerley03} simulation show very similar 
results when compared to FPEOS. Only small differences in temperature profile can be seen between 
the FPEOS and {\sl Kerley03} simulations, both of which are in remarkable contrast to the {\sl SESAME} case.  
Figure ~\ref{omega}(c) show the density profile at the peak compression, in which 
the predicted peak density ($\rho_p \simeq 210$ g/cm$^3$) is $\sim 25$\% lower 
according to FPEOS and {\sl Kerley03} compared 
to the {\sl SESAME} prediction ($\rho_p \simeq 260$ g/cm$^3$).
The history of areal density $\rho R$-evolution and 
neutron production were shown in Fig.~\ref{omega}(d). One sees that 
the peak $\rho R$ and neutron yield are also reduced by $\sim 10$\% - 20\% when the FPEOS and {\sl Kerley03}
are compared to the {\sl SESAME} predictions. 
The absolute neutron yield drops from $\sim 8.44 \times 10^{13}$ predicted by {\sl SESAME} to 
$\sim 6.91\times 10^{13}$ (FPEOS) and $\sim 6.93\times 10^{13}$ ({\sl Kerley03}).

Figure ~\ref{nif} shows the similar effects 
for the hydro-equivalent direct-drive NIF design with 1-MJ laser energy. 
The NIF target ($\phi = 2.954$-mm) consists of 27-$\mu m$ plastic ablator
and 170-$\mu m$ DT ice. The triple-picket drive pulse has a total duration 
of $\sim 11.4$ ns and a peak power of $\sim 240$-TW.  
We also found a decrease in $\rho _p$ and a slight temperature increase 
for the FPEOS and {\sl Kerley03} relative to {\sl SESAME} simulations near 
the end of the laser pulse ($t=9.2$ ns), shown by Fig. ~\ref{nif}(b).
The peak density at the stagnation
dropped from 481 ({\sl SESAME}) to $\sim 445$ g/cm$^3$ (FPEOS/{\sl Kerley03}), which is
indicated by Fig.~\ref{nif}(c). The resulting $\rho R$ and neutron yield
as a function of time is plotted in Fig. \ref{nif}(d). 
The yield dropped from the {\sl SESAME} value of $Y \simeq 1.75 \times 
10^{19}$ to $Y \simeq 1.57 \times 10^{19}$ (FPEOS) and $Y \simeq 1.55 \times 
10^{19}$ ({\sl Kerley03}).  Consequently,
the energy gain decreased from 49.1 ({\sl SESAME}) to 44.2 (FPEOS) and 
43.8 ({\sl Kerley03}). It is noted that the $\sim 11$\% gain reduction for 
this design is much modest than the $1.5$-MJ NIF design discussed in Ref. [\cite{FPEOS_PRL}]
in which more than $\sim 20$\% gain difference has been seen between FPEOS 
and {\sl SESAME} simulations. This is attributed to the different density-temperature
trajectories that the two designed implosions undergo, in which the EOS variations 
among FPEOS, {\sl SESAME} and {\sl Kerley03} are different.

\begin{figure}
\rotatebox{0}{\includegraphics[scale=1.0]{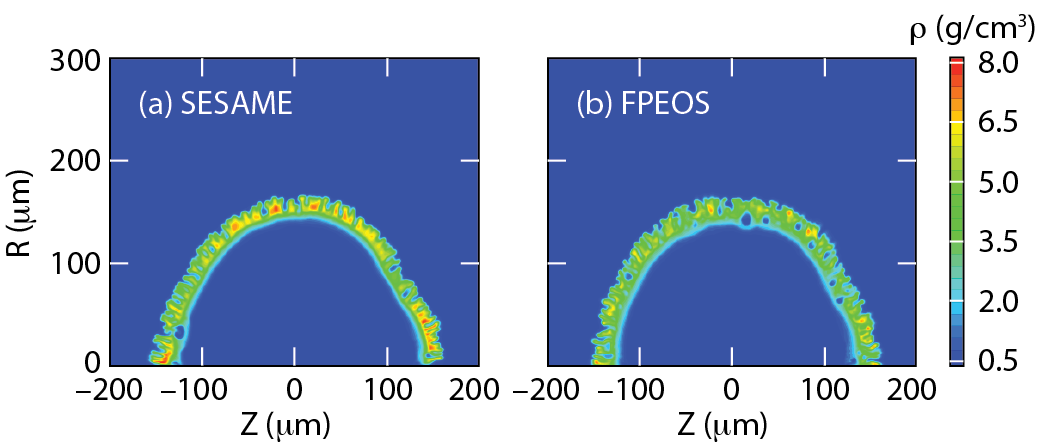}}
\caption{(Color online) The density contour plots at t=3.85 ns from two-dimensional DRACO simulations of 
the OMEGA cryogenic DT-implosion shown in Fig.~\ref{omega}, respectively using 
the {\sl SESAME} (a) and the FPEOS
(b) for the fuel DT. It is noted that the various perturbation sources have been included up to 
mode $\sl l =150$.
}
\label{Draco}
\end{figure}

Finally, we discuss the implications of the coupling and degeneracy effects in FPEOS to ICF target
performance beyond the 1D physics studied above. As we knew that various perturbations seeded 
by target roughness and lasers can grow {\sl via} the Rayleigh-Taylor (RT) instability \cite{RTs} 
during the shell acceleration/deceleration phases in ICF implosions, it is important to properly simulate 
the RT growth of fusion fuel for understanding target performance (compression and neutron yields)
\cite{my_POP_2009, my_POP_2010}. 
Since the RT growth depends on the compressibility of materials, the accurate 
equation-of-state of deuterium is essential to ICF designs. As an example, we    
have used our two-dimensional radiative hydro-code {\sl DRACO} to simulate 
the cryogenic DT implosion on OMEGA [discussed in Fig.~\ref{omega}].
The various perturbation sources, including the target offset, ice roughness, 
and laser irradiation non-uniformities measured from experiments, have been 
taken into account up to a maximum mode of $l=150$.
We have compared the FPEOS and {\sl SESAME} simulation results in 
Fig.~\ref{Draco} for $t=3.85$~ns near the end of acceleration, 
in which the density contours are plotted in the 
$ZR$-plane (azimuthal symmetry with respect to the Z-axis is assumed). 
Visible differences in the DT shell density can even be seen by eye from
Figs.~\ref{Draco}(a) and (b). The FPEOS simulation resulted in more ``holes" 
and density modulations along the shell than the {\sl SESAME} case.

\begin{figure}
\rotatebox{0}{\includegraphics[scale=1.0]{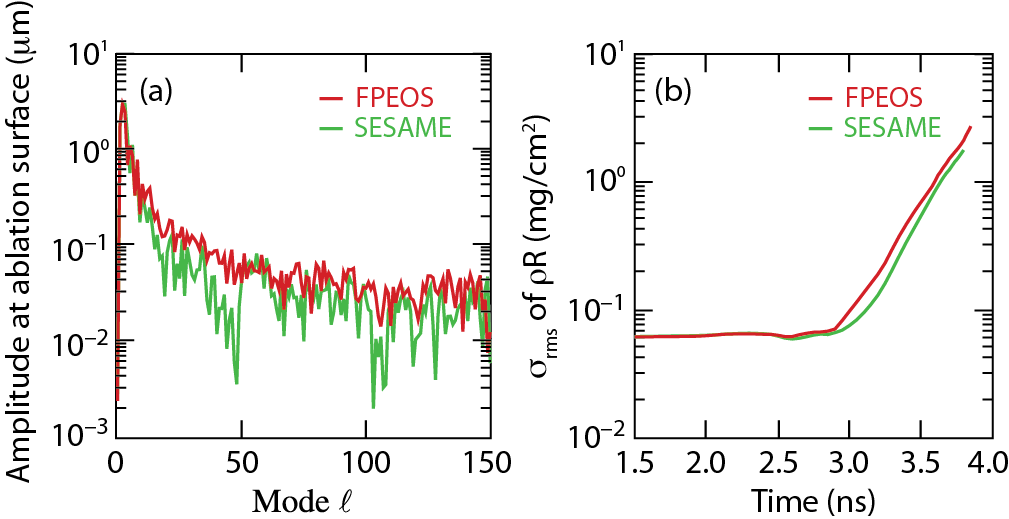}}
\caption{(Color online) (a) The modal spectrum of perturbation amplitude at the ablation surface for 
$t \simeq 3.0$ ns (the beginning of acceleration); (b) The perturbation growth in $\rho R$ as 
a function of time, analyzed from the 2D-DRACO simulations shown in Fig.~\ref{Draco}. 
}
\label{modes}
\end{figure}

To further analyze the perturbation amplitudes, we have decomposed the 
ablation-surface modulations into a modal spectrum that is shown in
Fig.~\ref{modes}(a), at the start of shell acceleration (t=3.0 ns).
We find that the FPEOS predicted larger amplitudes than the {\sl SESAME} 
case almost over the entire modal range. As the deuterium Hugoniot was shown 
in Ref. [\cite{FPEOS_PRL}], the FPEOS 
predicted softer deuterium compared to {\sl SESAME} for pressures below 
$\sim 2$ Mbar. Thus, the softer deuterium can be more easily ``imprinted'' by
the perturbations brought in {\sl via} the series of shocks. 
This results in larger perturbation amplitudes in FPEOS  than 
{\sl SESAME} simulations.
The Rayleigh-Taylor instability further amplifies these 
perturbations during the shell acceleration. As indicated by Fig.~\ref{modes}(b),
the $\sigma _{rms}$ of fuel $\rho R$ modulation increases to a few mg/cm$^2$ at the end of 
the laser pulse. These perturbations penetrated into the inner surface of the 
DT-shell will become the seeds for further RT growth during the shell's 
deceleration phase. They eventually distort the hot-spot temperature and density, thereby
reducing the neutron production. At the end, we found that
the {\sl SESAME} simulation resulted in a neutron-averaged ion 
temperature of $\left<T_i\right>=2.6$~keV 
and a neutron yield of $Y=5.2\times 10^{12}$; while due to 
the larger perturbations predicted 
the FPEOS simulation has given an $\left<T_i\right>=2.3$~keV 
and neutron yield of $Y=3.7\times 10^{12}$, which is more close to experimental
observations of $\left<T_i\right>=1.8 \pm 0.5$~keV and $Y=1.9\times 10^{12}$.

\section{Summary}

In conclusion, we have derived a first-principles equation of state
table of deuterium for ICF applications from PIMC calculations. The
derived FPEOS table covers the whole plasma density and temperature
conditions in low-adiabat ICF implosions. In comparison with the
chemical model based {\sl SESAME} table, the FPEOS table show
significant difference in internal energy and pressure for coupled and
degenerate plasma conditions; while the recently improved {\sl
  Kerley03} table exhibited fewer and smaller discrepancies when
compared to the FPEOS predictions temperature higher than $\sim
10$-eV.  Although subtle differences at lower temperatures ($T<10$ eV)
and moderate densities ($1 - 10$ \gcc) have been
identified and an artificial pressure cusp still exists in the {\sl
  Kerley03} table, radiation hydro-simulations of cryogenic ICF
implosions using the FPEOS and {\sl Kerley03} tables have given
similar peak density, areal density $\rho R$, and neutron yield, which
are remarkably different from the {\sl SESAME} simulations.  Both the
FPEOS and the {\sl Kerley03} predicted $\sim 25$\% less peak density,
$\sim 10$\% smaller $\rho R$, and $\sim 10$\%-20\% less neutron yield,
when compared to the {\sl SESAME} case.  Two-dimensional simulations
further demonstrated the significant differences in target performance
between the FPEOS and {\sl SESAME} simulations.  In general, the FPEOS
simulations resulted in better agreement with experimental
observations in terms of ion temperature and neutron yield.  It is
also noted that the extreme conditions covered by the FPEOS table are
also important in astrophysics and planetary sciences, for example, to
model the evolution of stars~\cite{star} and to understand the
thermodynamical properties of stellar matter~\cite{astro}.

\begin{acknowledgments}
This work was supported by U.S. Department of Energy Office of Inertial 
Confinement Fusion under Cooperative Agreement No. No. DE-FC52-08NA28302, 
the University of Rochester, and New York State Energy Research and Development Authority.
SXH would thank the support by the National Science Foundation 
under the NSF TeraGrid grant PHY110009 and this work was partially utilized the 
NICS' Kraken Supercomputer. BM acknowledges support from NSF and NASA.
\end{acknowledgments}

*{E-mail: shu@lle.rochester.edu}


\begin{thebibliography} {999}

\bibitem{ICF}J. Nuckolls, L. Wood, A. Thiessen, and G. Zimmerman, Nature (London)
{\bf 239}, 139 (1972); S. Atzeni and J. Meyer-ter-Vehn, {\sl The Physics of Inertial Fusion} 
(Clarendon  Press, Oxford, 2004).

\bibitem{DDI}R. L. McCrory {\sl et al.,} Phys. Plasmas {\bf 15}, 055503 (2008);
D.D. Meyerhofer et al., Nuclear Fusion {\bf 51}, 053010 (2011).

\bibitem{IDI} J. D. Lindl, Phys. Plasmas {\bf 2}, 3933 (1995).

\bibitem{Betti}R. Betti and C. Zhou, Phys. Plasmas {\bf 12}, 110702 (2005); 
R. Betti {\sl et al.}, Plasma Phys. Controlled Fusion {\bf 48}, B153 (2006). 

\bibitem{GoncharovPRL_2010}V. N. Goncharov, T. C. Sangster, T. R. Boehly, S. X. Hu, I. V.
Igumenshchev, F. J. Marshall, R. L. McCrory, D. D. Meyerhofer, P. B.
Radha, W. Seka, S. Skupsky, C. Stoeckl, D. T. Casey, J. A. Frenje, and R.
D. Petrasso, Phys. Rev. Lett. {\bf 104}, 165001 (2010).

\bibitem{NIF}E. M. Campbell and W. J. Hogan, Plasma Phys. Control. Fusion {\bf 41}, B39 (1999).

\bibitem{omega} T. R. Boehly {\sl et al.}, Opt. Commun. {\bf 133}, 495 (1997).

\bibitem{Hu_PRL_2008} S. X. Hu {\sl et al.,} \prl {\bf 100}, 185003 (2008).

\bibitem{Kerley1972}G. I. Kerley, Phys. Earth Planet. Inter. {\bf 6}, 78 (1972).

\bibitem{Kerley2003}G. I. Kerley, Sandia National Laboratory, 
Technical Report No. SAND2003-3613, 2003 (unpublished).

\bibitem{SaumonPRA1992}D. Saumon, G. Chabrier, \pra {\bf 46}, 2084 (1992).

\bibitem{Ross1998}M. Ross, \prb {\bf 58}, 669 (1998).

\bibitem{Rogers2001}F. J. Rogers, Contrib. Plasma Phys. {\bf 41}, 179 (2001).

\bibitem{Juranek2002}H. Juranek, R. Redmer, Y. Rosenfeld, J. Chem. Phys. {\bf 117}, 1768 (2002).

\bibitem{PierleoniPRL1994}C. Pierleoni {\sl et al.}, \prl {\bf 73}, 2145 (1994).

\bibitem{MagroPRL1996}W. R. Magro {\sl et al.}, \prl {\bf 76}, 1240 (1996).

\bibitem{MilitzerPRL2000}B. Militzer, D.M. Ceperley, \prl {\bf 85}, 1890 (2000).

\bibitem{MilitzerPRL2001}B. Militzer, D. M. Ceperley , J. D. Kress, J. D. Johnson, L. A. Collins, and S. Mazevet, 
 \prl {\bf 87},  275502 (2001). 

\bibitem{LILAC}J. Delettrez {\sl et al.,} Phys. Rev. A {\bf 36}, 3926 (1987).

\bibitem{FPEOS_PRL}S. X. Hu, B. Militzer, V. N. Goncharov,  and S. Skupsky,  Phys. Rev. Lett. {\bf 104}, 235003 (2010).

\bibitem{relax-rates}M. S. Murillo and M. W. C. Dharma-wardana, \prl {\bf 100}, 205005 (2008);
B. Jeon {\sl et al.,} \pre {\bf 78}, 036403 (2008); G. Dimonte and J. Daligault, \prl
 {\bf 101}, 135001 (2008); J. N. Glosli {\sl et al.,} \pre {\bf 78}, 025401(R) (2008);
L. X. Benedict {\sl et al.,}  \prl {\bf 102}, 205004 (2009);
B. Xu and S. X. Hu, Phys. Rev. E {\bf 84}, 016408 (2011). 

\bibitem{French_PRL_2008}V. Recoules,  F. Lambert, A. Decoster, B. Canaud, and J. Clerouin 
 {\sl et al.,} \prl {\bf 102}, 075002 (2009).

\bibitem{Polluk2004}E. L. Pollock, B. Militzer, \prl {\bf 92}, 021101 (2004).

\bibitem{Kress2010} J. D. Kress, J. S. Cohen, D. A. Horner, 
F. Lambert, and L. A. Collins, Phys. Rev. E {\bf 82}, 036404 (2010).

\bibitem{Silva1997}L. B. Da Silva {\sl et al.}, \prl {\bf 78}, 483 (1997).

\bibitem{Collins1998Science}G. W. Collins {\sl et al.,} Science {\bf 281}, 1178 (1998).

\bibitem{Collins1998POP}G. W. Collins {\sl et al.,} Phys. Plasmas {\bf 5}, 1864 (1998).

\bibitem{MostovychPRL}A. N. Mostovych {\sl et al.,} \prl {\bf 85}, 3870 (2000);
Phys. Plasmas {\bf 8}, 2281 (2001).

\bibitem{Boehly2004}T. R. Boehly {\sl et al.,} Phys. Plasmas {\bf 11}, L49 (2004).

\bibitem{Hicks2009}D. G. Hicks {\sl et al.,} \prb {\bf 79}, 014112 (2009).

\bibitem{Knudson2001}M. D. Knudson {\sl et al.,} \prl {\bf 87}, 225501 (2001);
{\sl ibid} {\bf 90}, 035505 (2003).

\bibitem{Knudson2004PRB}M. D. Knudson {\sl et al.,} \prb {\bf 69}, 144209 (2004).

\bibitem{Belov2002}S. I. Belov {\sl et al.,} JETP Lett. {\bf 76}, 433 (2002). 

\bibitem{Fortov2007}V. E. Fortov {\sl et al.,} \prl {\bf 99}, 185001 (2007).

\bibitem{LACollins1995}L. A. Collins {\sl et al.,} \pre {\bf 52}, 6202 (1995).

\bibitem{Lenosky2000}T. J. Lenosky {\sl et al.,} \prb {\bf 61}, 1 (2000).

\bibitem{GalliPRB2000}G. Galli {\sl et al.,} \prb {\bf 61}, 909 (2000).

\bibitem{CollinsPRB2001}L. A. Collins {\sl et al.,} \prb {\bf 63}, 184110 (2001).

\bibitem{Clerouin2001}J. Clerouin, J.F. Dufreche, \pre {\bf 64}, 066406 (2001).

\bibitem{DesjarlaisPRB2003}M. P. Desjarlais,  \prb {\bf 68}, 064204 (2003).

\bibitem{BonevPRB2004}S. A. Bonev, B. Militzer, G. Galli,  \prb {\bf 69}, 014101 (2004).

\bibitem{Collins_QMD} L. A. Collins (private communication).

\bibitem{OFMD}F. Lambert, J. Clerouin, and G. Zerah, Phys. Rev. E {\bf 73}, 016403 (2006);
F. Lambert, J. Clerouin, and S. Mazevet, Europhys. Lett. {\bf 75}, 681 (2006);
D. A. Horner, F. Lambert, J. D. Kress, and L. A. Collins, Phys. Rev. B {\bf 80}, 024305 (2009).

\bibitem{Mi06}B. Militzer, \prl {\bf 97}, 175501 (2006).

\bibitem{Milizer_He_2009}B. Militzer, Phys. Rev. B {\bf 79}, 155105 (2009).

\bibitem{MiltzerPhD}B. Militzer, PhD Thesis (University of Illinois, 2000).

\bibitem{CeperleyRMP1995}D. M. Ceperley, Rev. Mod. Phys. {\bf 67}, 279 (1995).

\bibitem{Ce91}D. M. Ceperley, J. Stat. Phys. {\bf 63}, 1237 (1991).

\bibitem{Ce96}D. M. Ceperley, in {\sl Monte Carlo and Molecular Dynamics of Condensed Matter Systems}, 
edited by K. Binder and G. Ciccotti (Editrice Compositori, Bologna, Italy, 1996).


\bibitem{MP00}B. Militzer and E. L. Pollock, \pre {\bf 61}, 3470 (2000).

\bibitem{Mi99}B. Militzer, W. Magro, D. M. Ceperley, Contrib. Plasma Phys. {\bf 39}, 151 (1999).

\bibitem{DH-model}P. Debye and E. H\"uckel, Phys. Z. {\bf 24}, 185 (1923).

\bibitem{draco}D. Keller, T. J. B. Collins, J.
A. Delettrez, P. W. McKenty, P. B. Radha, B. Whitney, and G. A. Moses,
Bull. Am. Phys. Soc. {\bf 44}, 37 (1999);
P. B. Radha, T. J. B. Collins, J. A. Delettrez, Y. Elbaz, R. Epstein, V. Yu.
Glebov, V. N. Goncharov, R. L. Keck, J. P. Knauer, J. A. Marozas, F. J.
Marshall, R. L. McCrory, P. W. McKenty, D. D. Meyerhofer, S. P. Regan,
T. C. Sangster, W. Seka, D. Shvarts, S. Skupsky, Y. Srebro, and C.
Stoeckl, Phys. Plasmas {\bf 12}, 056307 (2005);
S. X. Hu, V. A. Smalyuk, V. N. Goncharov, S. Skupsky, T. C. Sangster, D.
D. Meyerhofer, and D. Shvarts, Phys. Rev. Lett. {\bf 101}, 055002 (2008).

\bibitem{shock-timing1} T. R. Boehly {\sl et al.,} Phys. Plasmas {\bf 16}, 056302 (2009).

\bibitem{shock-timing2}T. R. Boehly, V. N. Goncharov, W. Seka, M. A. Barrios, P. M. Celliers, D. G. Hicks, G. W. Collins, S. X. Hu,
 J. A. Marozas, and D. D. Meyerhofer, Phys. Rev. Lett. {\bf 106}, 195005 (2011). 

\bibitem{qeos}R. More, K. H. Warren, D. A. Young, and G. Zimmermann, Phys. Fluids {\bf 31}, 3059 (1988).

\bibitem{RTs}B. A. Remington, S. V. Weber, M. M. Marinak, S. W. Haan, J. D.
Kilkenny, R. Wallace, and G. Dimonte, Phys. Rev. Lett. {\bf 73}, 545 (1994); H. Azechi,
T. Sakaiya, S. Fujioka, Y. Tamari, K. Otani, K. Shigemori, M. Nakai, H.
Shiraga, N. Miyanaga, and K. Mima, {\sl ibid.}  {\bf 98}, 045002 (2007);
V. A. Smalyuk, S. X. Hu, V. N. Goncharov, D. D. Meyerhofer, T. C.
Sangster, D. Shvarts, C. Stoeckl, B. Yaakobi, J. A. Frenje, and R. D.
Petrasso, {\sl ibid.}  {\bf 101}, 025002 (2008);
V. A. Smalyuk, S. X. Hu, V. N. Goncharov, D. D. Meyerhofer, T. C.
Sangster, C. Stoeckl, and B. Yaakobi, Phys. Plasmas {\bf 15}, 082703 (2008);
V. A. Smalyuk, S. X. Hu, J. D. Hager, J. A. Delettrez, D. D. Meyerhofer,
T. C. Sangster, and D. Shvarts, Phys. Rev. Lett. {\bf 103}, 105001 (2009);
V. A. Smalyuk, S. X. Hu, J. D. Hager, J. A. Delettrez, D. D. Meyerhofer,
T. C. Sangster, and D. Shvarts, Phys. Plasmas {\bf 16}, 112701 (2009). 

\bibitem{my_POP_2009}S. X. Hu {\sl et al.}, Phys. Plasmas {\bf 16}, 112706 (2009).

\bibitem{my_POP_2010}S. X. Hu {\sl et al.}, Phys. Plasmas {\bf 17}, 102706 (2010).

\bibitem{star} F. J. Rogers and A. Nayfonov, Astrophys. J. {\bf 576}, 1064 (2002).

\bibitem{astro} W. Stolzmann and T. Bl\"ocker, Astron. Astrophys. {\bf 361}, 1152 (2000).


\end{thebibliography}
\end{document}